\documentclass{aastex63}

\def\apjl{{\it Astrophys. J.} Lett~}

\def\msun{M_\odot}

\def\aap{{\it A\&A~}}

\shorttitle{Deep Learning Gravitational Wave Forecasting for Eccentric Binary Mergers}
\shortauthors{Wei Wei et al.}
\graphicspath{{./}{fig/}}

\begin{document}

\title{Deep Learning with Quantized Neural Networks for Gravitational Wave Forecasting \\ of Eccentric Compact Binary Coalescence}

\correspondingauthor{Wei Wei}
\email{weiw2@illinois.edu}

\author[0000-0003-1187-2253]{Wei Wei}
\affiliation{National Center for Supercomputing Applications, University of Illinois at Urbana-Champaign Urbana, Illinois 61801, USA}
\affiliation{NCSA Center for Artificial Intelligence Innovation, University of Illinois at Urbana-Champaign, Urbana, Illinois 61801, USA}
\affiliation{Department of Physics, University of Illinois at Urbana-Champaign, Urbana, Illinois 61801, USA}

\author[0000-0002-9682-3604]{E. A. Huerta}
\affiliation{Data Science and Learning Division, Argonne National Laboratory, Lemont, Illinois 60439, USA}
\affiliation{University of Chicago, Chicago, Illinois 60637, USA}
\affiliation{National Center for Supercomputing Applications, University of Illinois at Urbana-Champaign Urbana, Illinois 61801, USA}
\affiliation{Department of Physics, University of Illinois at Urbana-Champaign, Urbana, Illinois 61801, USA}
\affiliation{Department of Astronomy, University of Illinois at Urbana-Champaign, Urbana, Illinois 61801, USA}

\author[0000-0001-5461-636X]{Mengshen Yun}
\affiliation{National Center for Supercomputing Applications, University of Illinois at Urbana-Champaign Urbana, Illinois 61801, USA}
\affiliation{NCSA Center for Artificial Intelligence Innovation, University of Illinois at Urbana-Champaign, Urbana, Illinois 61801, USA}
\affiliation{Department of Electrical and Computer Engineering, University of Illinois at Urbana-Champaign, Urbana, Illinois 61801, USA}

\author[0000-0002-1597-3281]{Nicholas Loutrel}
\affiliation{Department of Physics, Princeton University, Princeton, New Jersey 08544, USA}
\affiliation{Princeton Gravity Initiative, Princeton University, Princeton, New Jersey 08544, USA}

\author[0000-0003-0826-6164]{Md Arif Shaikh}
\affiliation{International Centre for Theoretical Sciences, Tata Institute of Fundamental Research, Bangalore 560089, India}

\author[0000-0001-5523-4603]{Prayush Kumar}
\affiliation{International Centre for Theoretical Sciences, Tata Institute of Fundamental Research, Bangalore 560089, India}
\affiliation{Cornell Center for Astrophysics and Planetary Science, Cornell
University, Ithaca, New York 14853, USA}

\author[0000-0003-1424-6178]{Roland Haas}
\affiliation{National Center for Supercomputing Applications, University of Illinois at Urbana-Champaign Urbana, Illinois 61801, USA}

\author[0000-0002-9336-4756]{Volodymyr Kindratenko}
\affiliation{National Center for Supercomputing Applications, University of Illinois at Urbana-Champaign Urbana, Illinois 61801, USA}
\affiliation{NCSA Center for Artificial Intelligence Innovation, University of Illinois at Urbana-Champaign, Urbana, Illinois 61801, USA}
\affiliation{Department of Electrical and Computer Engineering, University of Illinois at Urbana-Champaign, Urbana, Illinois 61801, USA}
\affiliation{Department of Computer Science, University of Illinois at Urbana-Champaign, Urbana, Illinois 61801, USA}



\begin{abstract}

\noindent We present the first application of deep 
learning forecasting for binary neutron stars, neutron 
star-black hole systems, 
and binary black hole mergers that span an eccentricity 
range \(e\leq0.9\). We train neural networks that describe 
these astrophysical populations, and then test their performance
by injecting simulated eccentric signals in advanced LIGO noise available at 
the \texttt{Gravitational Wave Open Science Center} 
to: 1) quantify how fast neural networks 
identify these signals before the binary components merge; 2) 
quantify how accurately neural networks estimate the time to 
merger once gravitational waves are identified; and 3) estimate 
the time-dependent sky localization of these events from 
early detection to merger. Our findings show that deep learning 
can identify eccentric signals from a few seconds (for binary black 
holes) up to tens of seconds (for binary neutron stars) 
prior to merger. A quantized version of our neural networks 
achieves 4x reduction in model size, and up to 2.5x inference 
speed up. These novel algorithms may be used to facilitate 
time-sensitive multi-messenger astrophysics observations of 
compact binaries in dense stellar environments.
\end{abstract}

\keywords{Gravitational Waves --- 
Deep Learning --- Forecasting --- Eccentric Mergers --- Advanced LIGO}


\section{Introduction}
\label{intro}

Multi-Messenger observations that combine the gravitational 
and electromagnetic spectra~\citep{bnsdet:2017,mma:2017arXiv,Hurtley1551,grb:2017ApJ,kiloGW170817:2017,Mooley:2017enz,2017Natur55171T} have provided 
revolutionary insights about 
the nature of gravity, the engines that power short gamma ray bursts, 
cosmology and fundamental physics~\citep{Soares_Santos_2019,Abbott,Hubble_from_GW,Maya:2018F,Berti:2018GReGr,LIGOScientific:2019fpa}. 
These remarkable discoveries provide a glimpse of what 
Multi-Messenger Astrophysics may accomplish once gravitational 
wave detectors reach their design sensitivity, and they work 
in unison with electromagnetic and astro-particle observatories to 
observe the transient universe with 
unprecedented precision~\citep{LIGOScientific:2019gag,mma_nature_revs,2019NatRP...1..585M}. To realize these goals, however, 
there is an urgent need to develop signal processing tools 
and computing frameworks that turn computational grand 
challenges in the big-data era into unique opportunities 
to enable new modes of data-driven 
discovery~\citep{2019NatRP...1..600H,mma_nature_revs,huerta_zhao_chapter}.

In the realm of gravitational wave observations, recent 
developments include the production of early warning systems 
to forecast the merger of Multi-Messenger sources using 
template-matching methods in the context of simulated advanced LIGO 
noise~\citep{2012ApJ...748..136C,Sachdev:2020lfd}, and third generation 
ground-based detectors~\citep{2020ApJ...902L..29N}. Deep 
learning has  
emerged as a powerful tool to process data at scale, 
with similar sensitivity of traditional 
algorithms, but at a fraction of their computational cost. 
Deep learning methods have evolved rapidly in 
gravitational wave astrophysics, ranging from the first 
algorithms that were proposed to enable real-time gravitational wave 
detection with advanced 
LIGO 
noise~\citep{geo_hue:NN_Sim_LIGO,George:2017vlv,geo_hue:NN_Real_LIGO}, 
to the production of sophisticated neural networks that 
span the same signal manifold of traditional low-latency 
pipelines, process hundreds of hours of advanced LIGO noise 
faster than real-time with just a handful of GPUs, 
and identify real events with a minimal number of false 
positives in real advanced 
LIGO noise~\citep{Wei_Khan_Hue:NN_Scale,huerta_ai_hpc_foster}. Deep 
learning has also been 
used to forecast the merger of quasi-circular binary neutron stars and black hole-neutron stars systems in real advanced LIGO 
noise~\citep{Wei_forecast}.

To date, neural networks have been developed in the 
context of quasi-circular gravitational wave sources, even 
though some of these models have been used to explore the 
detection of eccentric binary black hole mergers~\citep{Adam:2018prd}. 
In this paper 
we introduce the first class of neural networks 
that are trained and tested with waveforms that 
describe binary neutron stars, neutron star-black hole mergers, and 
binary black hole mergers that cover a broad eccentricity 
range, \(e\leq0.9\). These neural networks target a 
significantly much more 
challenging task than detection, since these models can predict the 
merger of eccentric signals embedded in advanced LIGO data 
from a few seconds up to two minutes before the merger event. 

This study is motivated by a number of considerations. 
For instance, it is well know that eccentric waveforms have a 
rich and complex morphology at early frequencies. Given that 
neural networks have been particularly successful at time-series 
processing and pattern identification~\citep{DeepLearning}, it 
is worth exploring whether we can we train neural networks 
to enable gravitational wave forecasting by leveraging the rich 
spectrum of frequencies that characterize eccentric signals 
in the context of advanced LIGO noise. If this analysis was 
indeed possible, we would like to quantify how fast 
deep learning may predict the merger of 
likely multi-messenger 
sources---neutron star mergers or stellar mass black 
hole-neutron star systems---or binary black holes 
that coalesce in dense stellar environments. This is the 
theme of this article.

We organize this article as follows. 
Section~\ref{sec:methods} describes our 
deep learning algorithms, the modeled waveforms and advanced LIGO noise 
used to train and test our models. We present our forecasting results in 
Section~\ref{sec:results}. We analyze network quantization results in Section~\ref{sec:quant}. We summarize our findings and discuss future activities in Section~\ref{sec:end}.


\section{Methods}
\label{sec:methods}

Here we describe the modeled waveforms and advanced LIGO noise used to 
create our neural networks, and how these 
models may be used to forecast the merger of compact binary systems 
in advanced LIGO noise. 

\subsection{Waveforms and advanced LIGO noise}
\label{sec:curation}

\noindent \textbf{Modeled waveforms} We used the waveform model introduced in~\citep{East:2012xq} to describe eccentric 
compact binary systems. Even though this is an admittedly 
approximate model, it provides a complete description of the 
systems' dynamics, from inspiral to ringdown, and more 
importantly, it enables the modeling of highly eccentric 
systems. The waveforms used in this study are produced at a sample rate 
of 16384Hz, and describe binary neutron stars, black hole-neutron star systems, and 
binary black hole mergers. For binary neutron stars, we consider systems with 
binary components \(m_{\{1,2\}}\in[1\msun, 3\msun]\), whereas for black hole-neutron star systems we considered \(m_{\textrm{BH}}\in[3\msun, 15\msun]\) and \(m_{\textrm{NS}}\in[1\msun, 3\msun]\) for the masses of the black hole and neutron star components, respectively. 



The waveforms in the training dataset cover the 
eccentricity range \(e\leq0.9\), and are 160s long. These waveforms are randomly split 
into training sets 
($12423$ waveforms for binary neutron stars, $15593$ for black hole-neutron star systems, and 10677 for binary black holes), and test sets ($3075$ waveforms for binary neutron stars, $3882$ for black hole-neutron star systems, and 2661 for binary black holes). 

\noindent \textbf{Advanced LIGO noise} We have used four 4096s-long 
advanced LIGO noise segments, sampled at 16384Hz, from the 
Hanford and Livingston detectors. These segments have GPS starting 
times $1186725888$, $1187151872$, 
$1187569664$, and $1186897920$. The first three segments 
are used for training, while the last is used for testing. All 
these open source data were obtained from the \texttt{Gravitational Wave Open Science Center}~\citep{Vallisneri:2014vxa}.

We rescaled and injected the
waveform datasets described above into real advanced LIGO noise, using both Livingston and Hanford data, to simulate eccentric 
mergers that span a broad range of signal-to-noise ratios (SNRs). We standardized the datasets 
for training by normalizing the standard deviation of the 
advanced LIGO strain data with signal injections.

\subsection{Spectrograms of Eccentric Compact Binary Mergers}

\begin{figure*}[!h]
    \centerline{
    \includegraphics[width=0.45\linewidth]{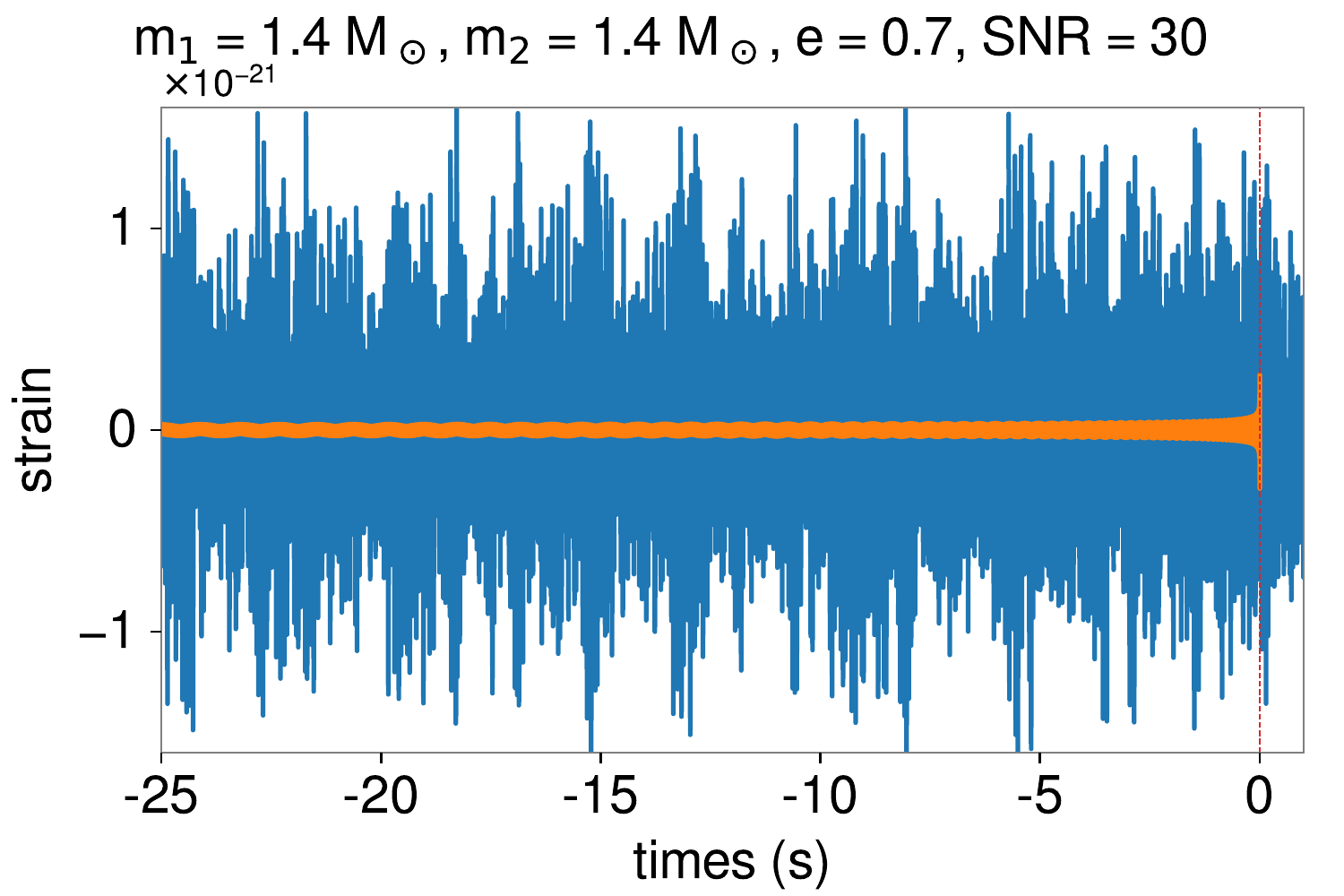}
    \includegraphics[width=0.45\linewidth]{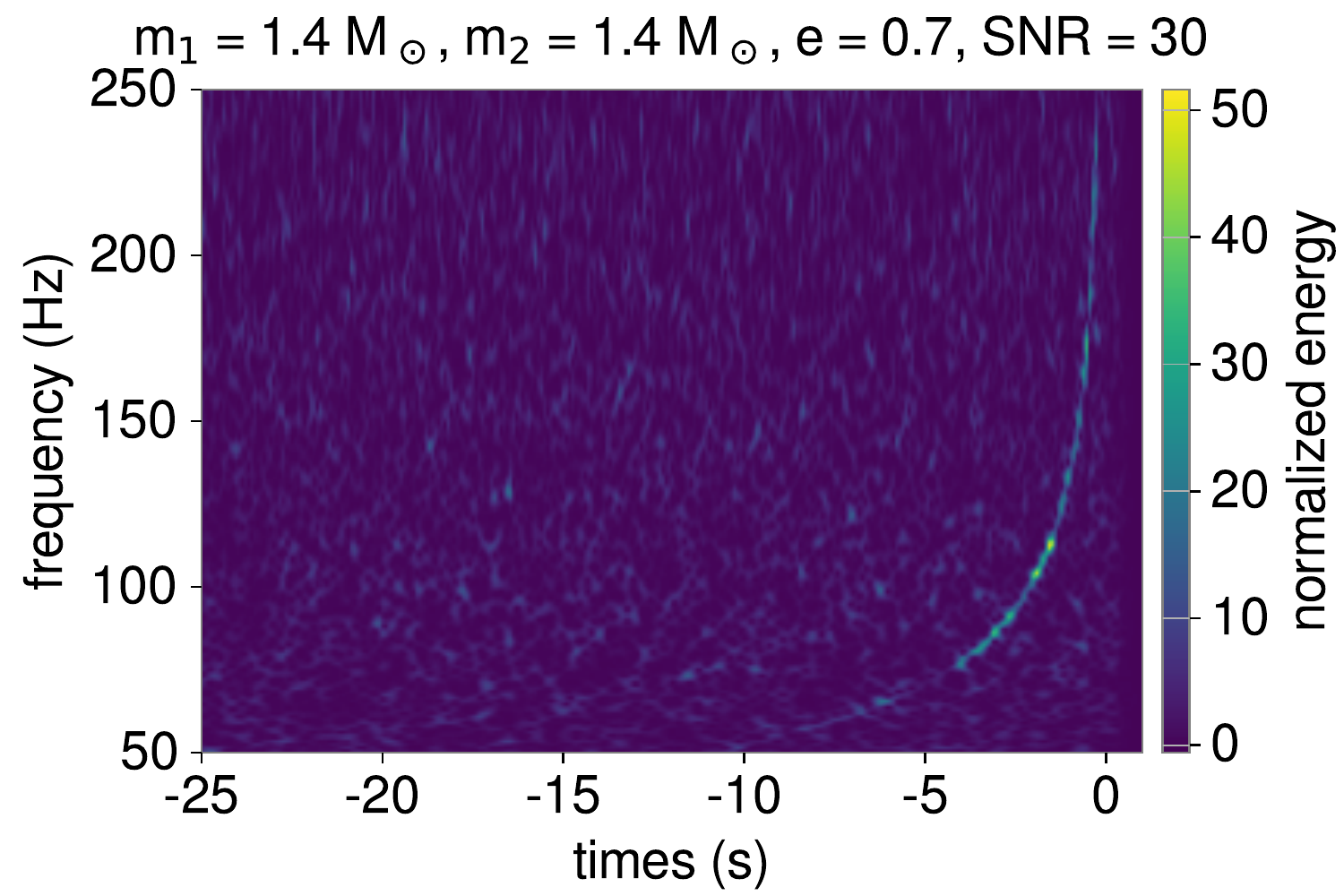}
    }
    \caption{Left panel: gravitational wave signal injected in advanced LIGO O2 noise that describes a binary neutron star with component masses \(m_{\{1,2\}}=1.4\msun\), eccentricity 
\(e=0.7\) measured 150 seconds before merger. Right panel: spectrum of frequencies of the eccentric waveform signal shown in the left panel.}
    \label{fig:ecc_spectrogram}
\end{figure*}

Eccentric binary neutron stars have long-duration inspiral stages with periodic spikes in time domain, as shown in the right panel of Figure~\ref{fig:ecc_spectrogram}. Their spectral decomposition provides a rich spectrum of frequencies, as shown in the right panel of Figure~\ref{fig:ecc_spectrogram}. We use the early inspiral-stage presence of the chirp patterns in the spectrograms to provide early warnings for imminent merger events, which may provide early warnings for potential electromagnetic follow-ups.

\subsection{Chirp-pattern recognition and merger-time estimation with deep learning}
\label{sec:patter_recognition}

\noindent \textbf{Early Detection} We use a deep neural network to identify inspiral-stage chirp-like signatures in spectrograms, and provide tens of seconds 
early warnings for a variety of compact binary systems.

Specifically, we use \texttt{ResNet-50}~\citep{resnet50} architecture pre-trained on \texttt{ImageNet}~\citep{cvpr:Deng}, implemented with \texttt{PyTorch}~\citep{NEURIPS2019_9015} for our pattern recognition tasks. We produce and stack together spectrograms from 8s-long advanced LIGO strain data to form an image of two channels with the first channel representing Livingston data, and the second channel representing Hanford data. We then pad a third channel with zeros so that the final images have three channels to match the \texttt{ResNet-50} architecture. The three-channel images made from spectrograms will be used as the input to \texttt{ResNet-50}. We also modify the last layer of the original \texttt{ResNet-50} design so that the output for input images is a single number from zero to one, indicating the probability for the presence of chirp signals in the input spectrogram image.

To ensure the output for an input image is a number in the range $[0,1]$, we used the \texttt{sigmoid} function defined as $\sigma(x) = 1/(1+\exp(-x))$. This function maps real numbers into the range of $[0,1]$, which can go on be interpreted as the probability for the presence of chirp signals. We choose a threshold of $0.8$ for the identification of signals, which means the time steps with output probabilities greater than $0.8$ indicate the presence of GW signals.

The spectrograms from 8s-long strain data are produced with a \texttt{blackman} window size of $16384$, and a step size of $1024$. The generated spectrograms have a size of $8193\times 113$ in the frequency and time domain, respectively. As a part of the pre-processing step, we also take the element-wise $\log$ transformation of the spectrograms to accentuate the chirp patterns. All the spectrograms are produced using the \texttt{spectrogram} function provided by \texttt{SciPy}.

Since the neural networks are trained to identify inspiral-stage chirp patterns in the spectrograms, we can truncate the spectrograms above $150$ Hz to reduce the size of the input images. We also remove the parts below $20$ Hz, which are dominated by low-frequency noise. Therefore, the actual size of the images used as the inputs to the trained neural network is $130\times 113$ with three channels as stated above.
\vspace{2mm}

\noindent \textbf{Merger Time Estimation} The neural network for 
merger time estimation is the same as the one used for chirp 
pattern detection, except that the last layer is a fully 
connected layer, and the network output corresponds to the 
estimated merger time. 

\subsection{Training strategy}
\label{sec:training}

\noindent \textbf{Early Detection} We separately trained three \texttt{ResNet-50}s on eccentric binary neutron star, black hole-neutron star and binary black hole datasets, as described in Section~\ref{sec:curation}. We followed the same training strategy for 
all cases. As mentioned above, we first injected clean waveforms into advanced LIGO noise data to simulate noisy signals with different SNRs. Then we produced spectrograms from those stimulated gravitational wave events to generate input images for \texttt{ResNet-50}. Finally, we trained \texttt{ResNet-50} using a batch size of $256$, and a learning rate of $10^{-4}$ with the ADAM~\citep{kingma2014adam} optimizer. The training and testing was done using 4 \texttt{NVIDIA V100} GPUs. For robust performance, we exposed the neural network to a variety of scenarios during the training. Specifically, $50\%$ of the input spectrogram images contain no gravitational waves, while $25\%$ have 
simulated signals only in either Livingston or Hanford data, 
while the remaining $25\%$ have waveform signals in both Livingston and Hanford data. Upon thoroughly testing the performance of our three 
separate neural networks, we found that the network trained on binary 
neutron star waveforms provided the best performance for early detection for all systems under consideration. We may understand 
this finding if we consider that forecasting depends critically on information the neural network extracts from the inspiral phase, 
and in the case of black hole-neutron star systems and binary black holes most of the power is concentrated in the vicinity of the 
merger. Thus, in what follows we present 
forecasting results using the neural network trained with binary neutron star waveforms for the three classes of binaries under 
consideration. 
\vspace{2mm}

\noindent \textbf{Merger Time Estimation}
For merger time estimation, the training process follows the 
same approach described above, except that the neural network 
is now trained to predict the merger time of the binary systems, 
and the input spectrogram images contain signals that merge 
at different times. We separately trained three neural networks on eccentric binary neutron star, neutron star-black hole and binary 
black hole datasets, as described in Section~\ref{sec:curation}.

\subsection{Sky Localization}
In addition to forecasting and quantifying the time to 
merger with deep learning, we have adapted an algorithm 
to rapidly estimate the sky localization 
area as a function of time, namely, from the time 
our neural networks identify a given signal until the 
merger event. The localization sky area is estimated 
via triangulation 
using a Fisher matrix based method, as described in~\citet{Fairhurst_2009, Fairhurst_2011}. The sky area 
may be estimated from the separation between the detectors, 
their individual effective bandwidth and the SNRs. The 
effective bandwidth, $\sigma_f$, of a detector is computed 
from the frequency moments as 

\begin{eqnarray}
    \sigma_f^2 &=& \bar{f^2} - \bar{f}^2\,, \quad \textrm{where}\\
    \bar{f^n} &=& 4 \int_0^\infty df\frac{f^n |\tilde{h}(f)|^2}{ S_n(f)}\,,
\end{eqnarray}

\noindent where $S_n$ is the noise spectral density, 
and $\tilde{h}$ is the frequency domain waveform. 
The effective bandwidth and the SNR, $\rho = \sqrt{\bar{f^0}}$, 
may be combined to compute the timing accuracy, $\sigma_t$, 
through the relation

\begin{equation}
    \sigma_t = \frac{1}{2\pi\rho\,\sigma_f}\,.
\end{equation}

\noindent The timing accuracy may be used to construct a posterior distribution of the source location~\citep{Fairhurst_2009, Fairhurst_2011}. In this study, we truncate our time domain signal at different times before merger and Fourier transform the truncated signal to get the frequency domain signal to compute the frequency moments which are then used to estimate the sky localization area as a function of time from early detection to merger. The results we 
present below assume a 3 detector network that encompasses the 
Hanford and Livingston LIGO detectors, and the Virgo detector. 
We have computed PSDs for each of these interferometers using 
the O2 noise segments described above.

\subsection{Quantized neural networks}
\label{sec:quantized_nets}

We have explored the use of quantized neural networks to enable 
forecasting at the edge in view of their compact size and power efficiency. To quantize our fully trained \texttt{ResNet-50} models, post-training static quantization is used to convert the weights and activations of the models from 32-bits to 8-bits representation. We utilize quantization tools provided by \texttt{PyTorch} to perform static quantization. First we define a \texttt{ResNet-50} model and insert a quantization layer at the beginning and a de-quantization layer at the end for handling the input and output tensors during inference. The weights of our trained \texttt{FP32} model are then loaded into the new model definition. Layer fusions are performed to fuse \texttt{Conv2D}, \texttt{BatchNorm}, and \texttt{ReLU} modules when possible to obtain better performance. 

We prepared our networks for inference by collecting statistics for each layer input, running calibration for the quantized model, and quantizing the trained weights into \texttt{INT8}. Asymmetric linear quantization is used here to scale and offset the values in activation tensors. The calibration step adjusts the scales and offsets to minimize accuracy loss due to quantization. The spectrogram images used for this step are randomly selected from the testing dataset, and only 30 images are required to fully calibrate the quantized parameters. After the networks have been quantized, we use \texttt{Intel Xeon E5-2620 CPU} with \texttt{AVX2} support as the backend for running the quantized networks. As we show below, our quantized networks have the same forecasting 
performance of regular neural networks, but are 4x more compact and 2.5x faster. These features promote them as ideal tools to enable 
gravitational wave forecasting at the edge.

\section{Results}
\label{sec:results}

We present results for three types of sources, 
binary neutron stars, black hole-neutron star systems, and 
binary black hole mergers. As mentioned above, our 
neural networks are used to search for patterns 
in spectrograms that characterize eccentric compact binary mergers. 
We used a sliding window of 8s, with a step size of 1s, that is applied to the spectrograms generated from strain data that are up to 160s long. The data within the sliding window are fed into 
the neural networks, and the neural networks output the probability for the existence of a gravitational wave signals in advanced 
LIGO noise.

\subsection{Eccentric Binary Neutron Stars}

Our first set of results comprise binary neutron stars with 
component masses \(1\msun \leq m_{\{1,2\}} \leq 2.1\msun\), and 
eccentricities \(e\leq0.9\). To test the performance of our 
neural networks, we prepared injections 
that sampled a broad range of inclinations, sky locations and 
SNRs. 

\begin{figure*}
\centerline{
\includegraphics[width=0.45\linewidth]{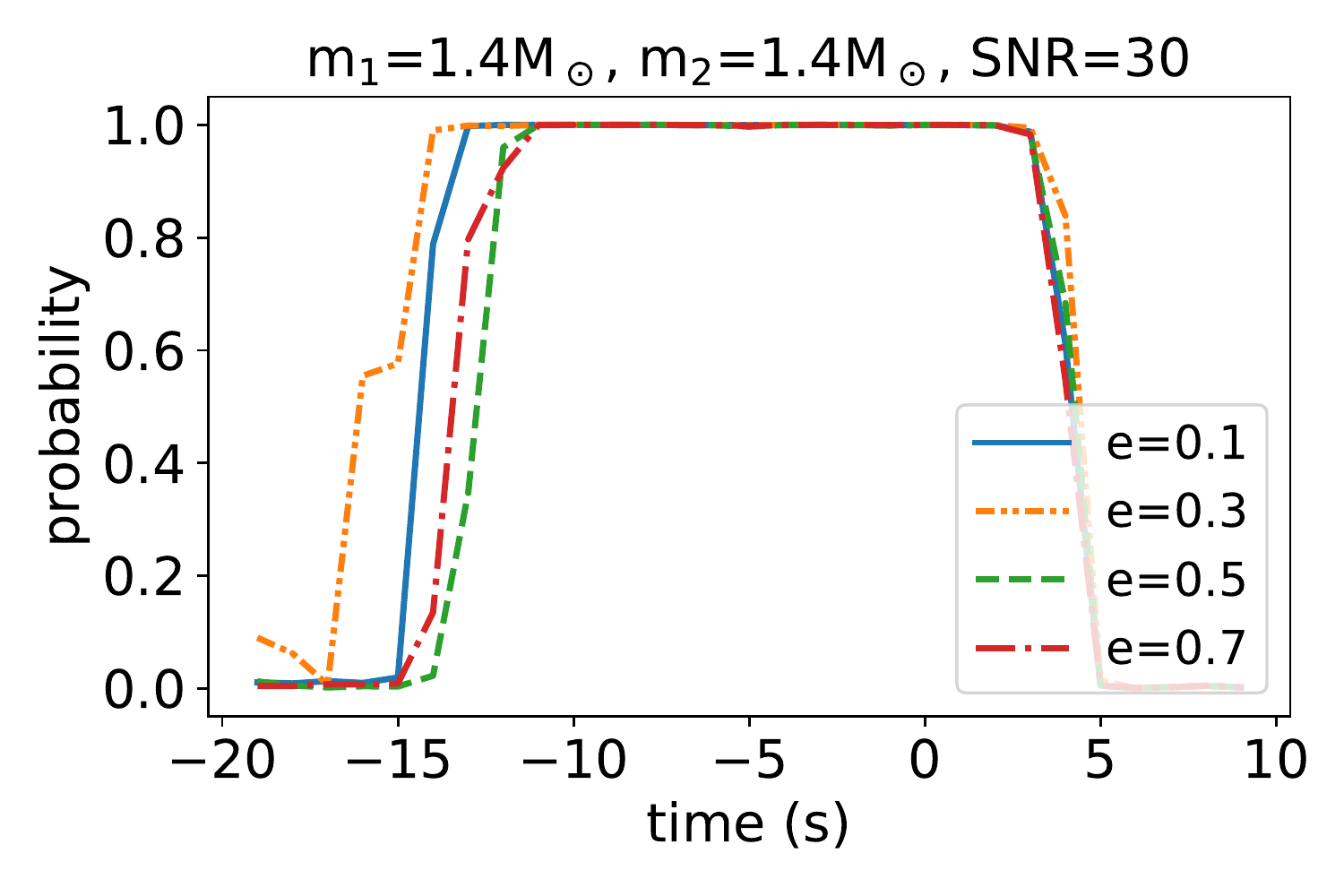}
\includegraphics[width=0.45\linewidth]{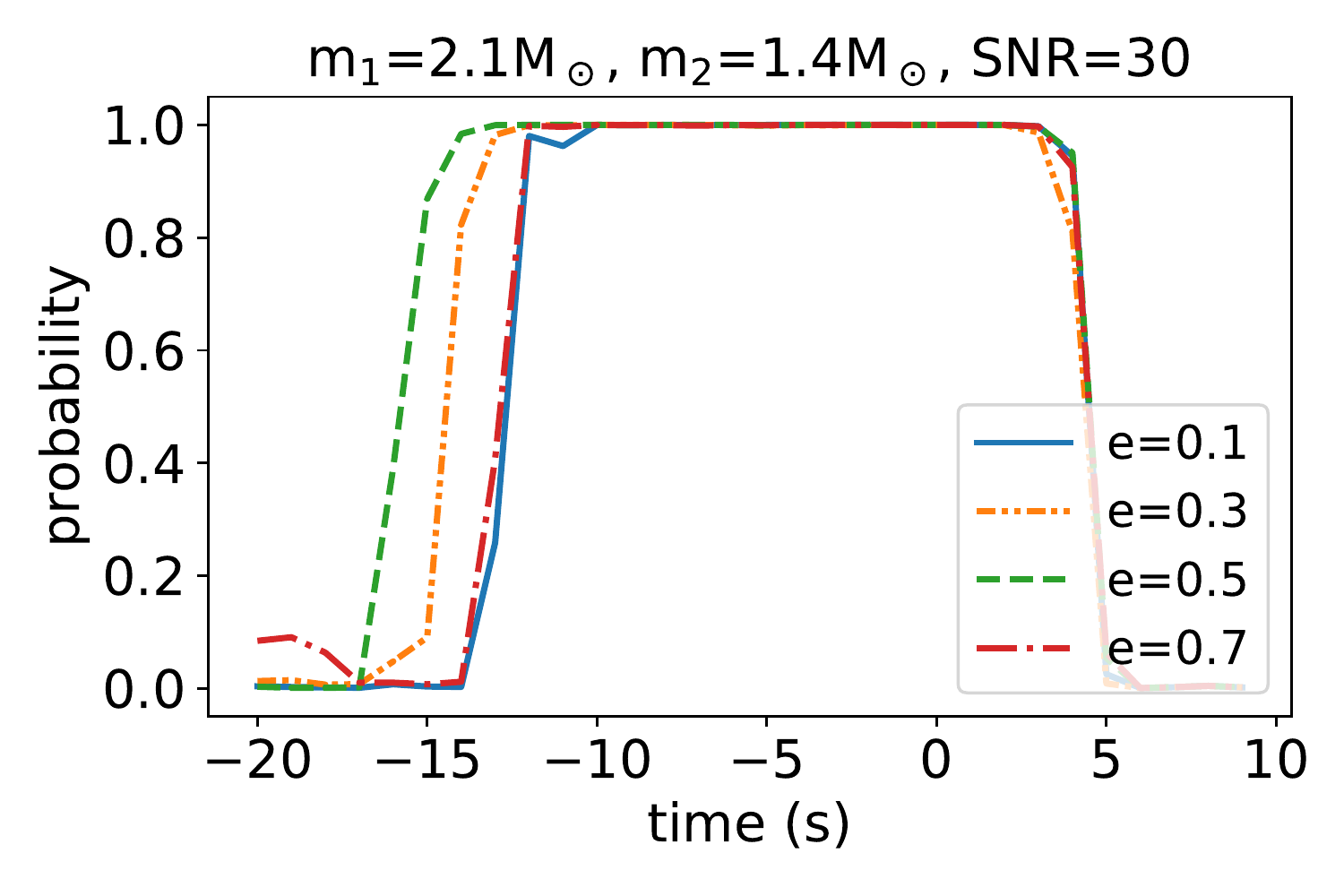}
}
\centerline{
\includegraphics[width=0.45\linewidth]{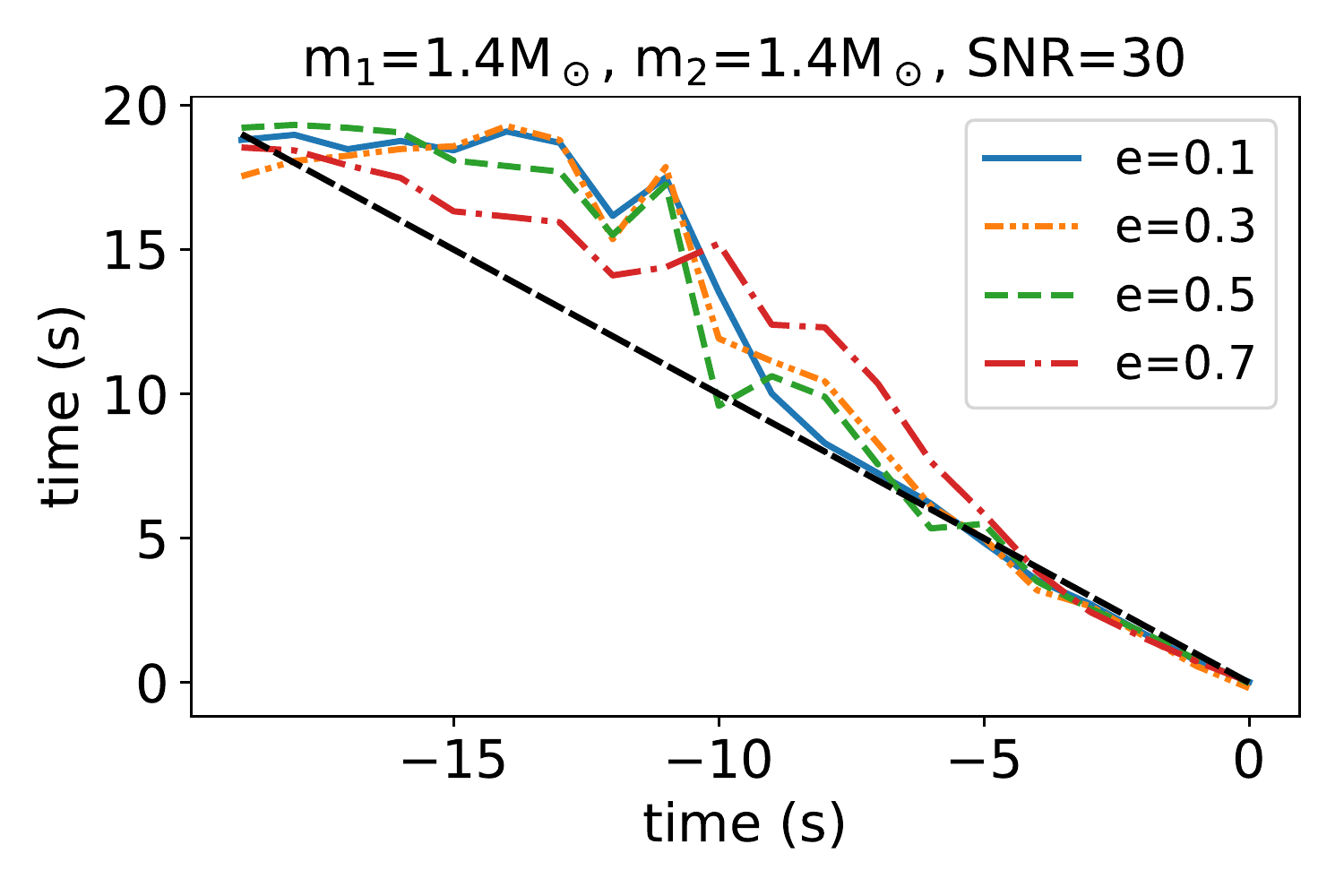}
\includegraphics[width=0.45\linewidth]{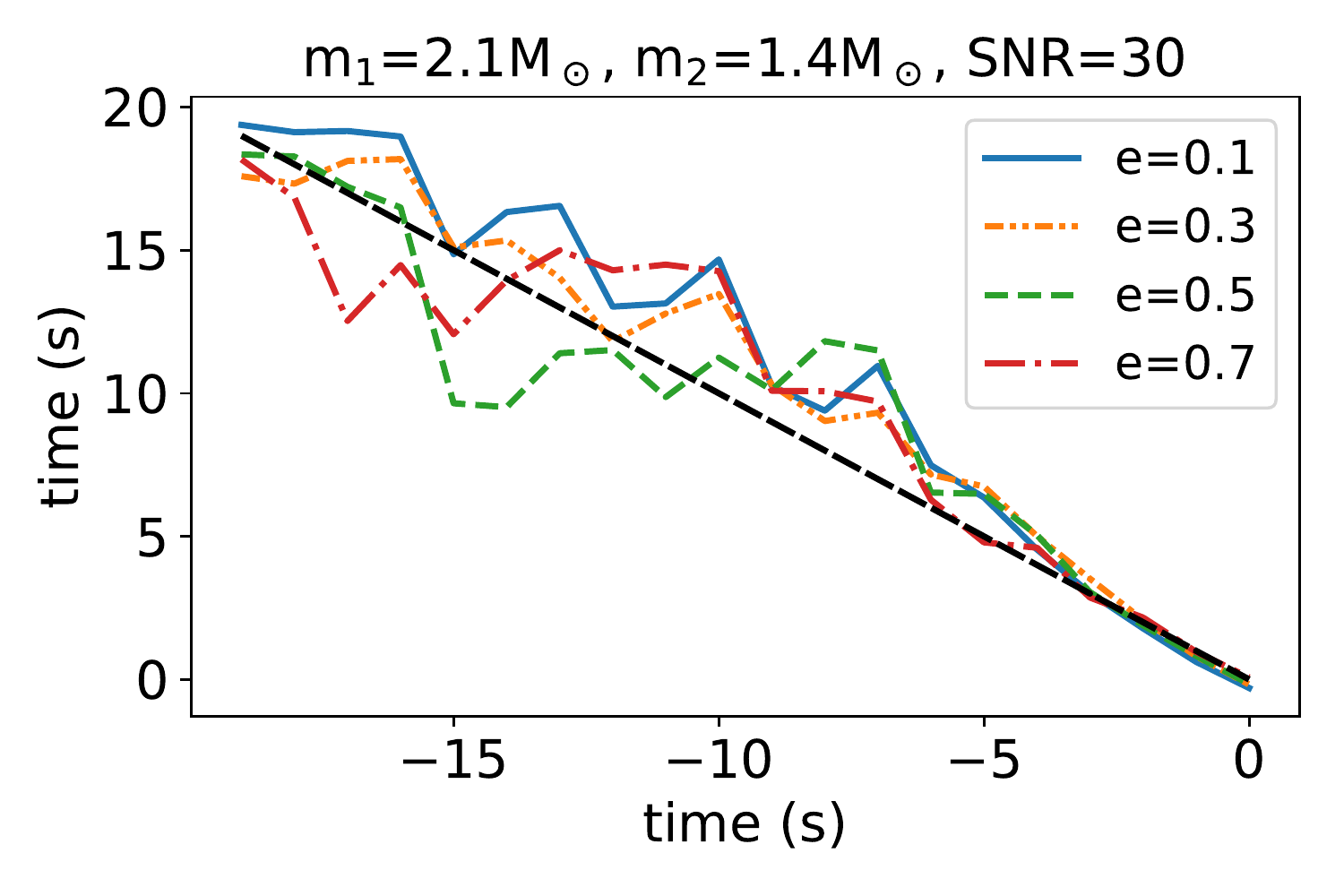}
}
\centerline{
\includegraphics[width=0.45\linewidth]{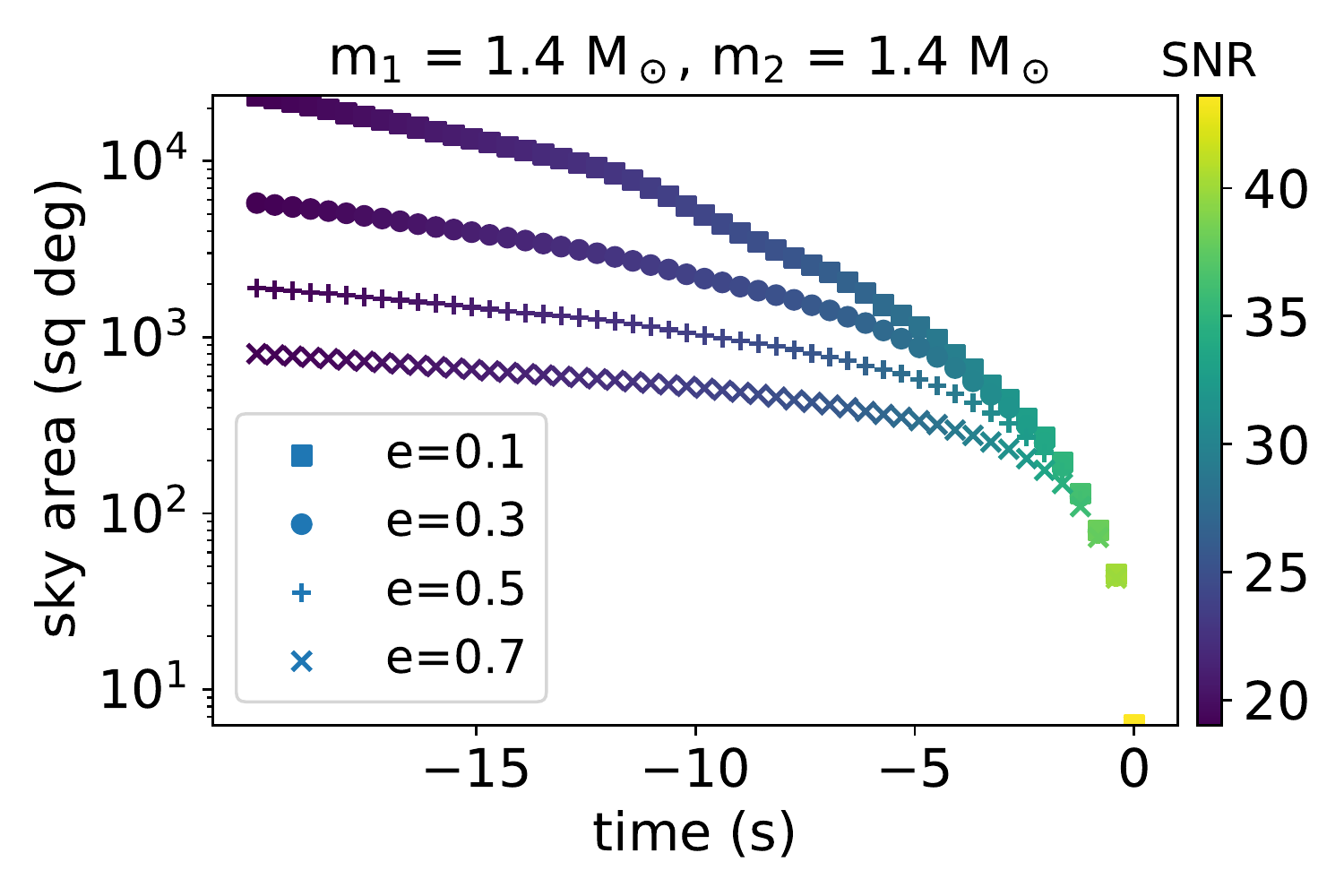}
\includegraphics[width=0.45\linewidth]{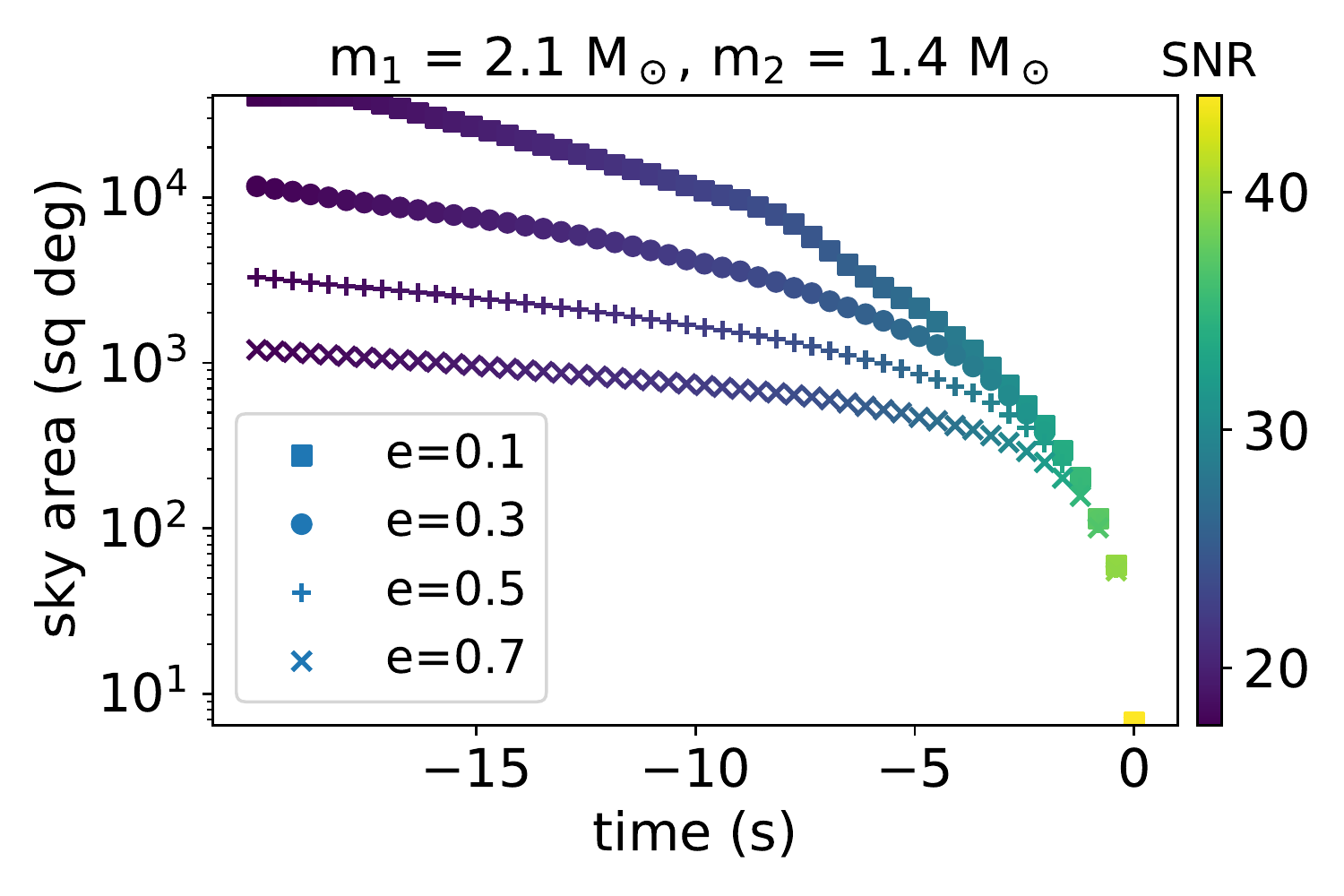}}
\caption{\textbf{Top panels} Neural networks identify injections of modeled binary neutron star waveforms in O2 LIGO data up to 15 seconds before merger. \textbf{Mid panels} Neural networks provide reliable estimates of time to merger 10 seconds ahead of the actual merger event.  \textbf{Bottom panels} Time-dependent sky localization of injected binary neutron star waveforms in O2 LIGO noise.}
\label{fig:ecc_bns}
\end{figure*}

In Figure~\ref{fig:ecc_bns} we present three sets of results 
for binary neutron stars with component masses \((m_1=m_2=1.4\msun)\), and \((m_1=2.1\msun, m_2=1.4\msun)\). In both cases we consider 
binaries with \(\textrm{SNR}=30\). The top panels in Figure~\ref{fig:ecc_bns} show that our neural networks identify 
these signals up to 15 seconds before their binary components 
coalesce. The mid panels show that our neural networks may 
provide a reliable estimate for the time to merger about 10 seconds 
before the binary components collide. Finally, the bottom panels in 
Figure~\ref{fig:ecc_bns} show that the sky localization is 
rather sensitivity to the eccentricity of the binary. We notice 
that sky localization improves by nearly two orders of 
magnitude from early detection up to merger for the most 
eccentric systems, and by three orders of magnitude for the least 
eccentric systems.

An extensive body of 
research in the literature argues strongly for the
modeling of eccentric binary neutron stars, and the new physics 
that may be learned by detecting these sources~\citep{East:2016PhRvD,East:2015PRDa,east:2012a,east:2012,Lehner:2014a,Gold:2012PG,PhysRevD.98.104005,2018PhRvD..98j4005C,Vick:2019cun,2018MNRAS.476..482V,PhysRevD.100.064023,PhysRevD.98.044007}. 
This new tool provides the means to enable such observations, 
and to even forecast when such objects may coalesce. If 
flybys during the inspiral evolution produce tidal disruptions 
with electromagnetic counterparts~\citep{Tsang:2013}, then deep learning 
forecasting may help associate these electromagnetic observations 
with the physics of eccentric neutron star systems.

\subsection{Eccentric Neutron Star-Black Hole Binaries}
\label{sec:nsbh}

As described above, we model black hole-neutron star binaries 
assuming systems with component masses 
\(m_{\textrm{BH}}\in[3\msun, 15\msun]\) and \(m_{\textrm{NS}}\in[1\msun, 3\msun]\), and eccentricities 
\(e\leq0.9\). As in the case of binary neutron stars, 
we prepared injections that sampled a broad range of 
inclinations, sky locations and SNRs.

\begin{figure*}
\centerline{
\includegraphics[width=0.45\linewidth]{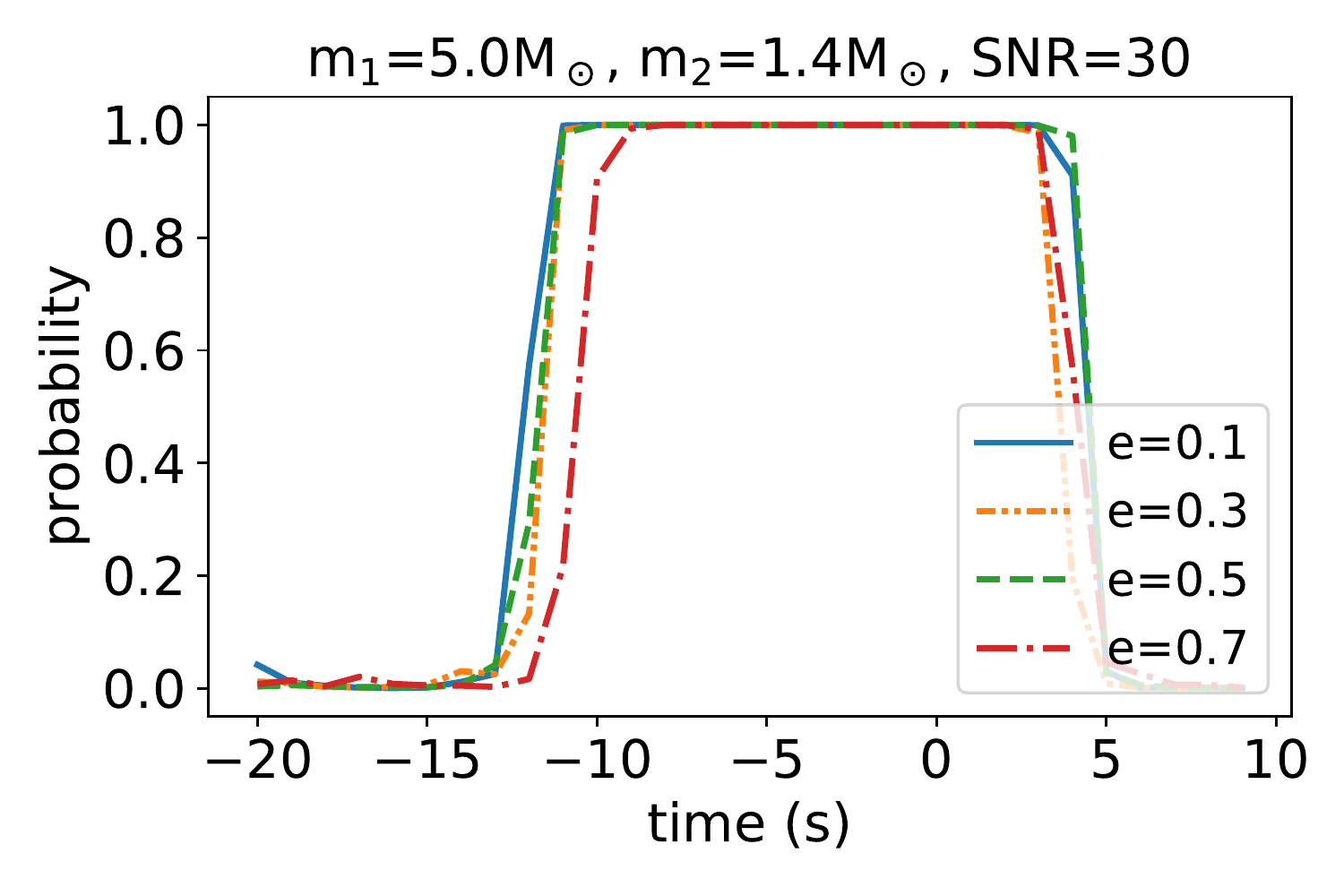}
\includegraphics[width=0.45\linewidth]{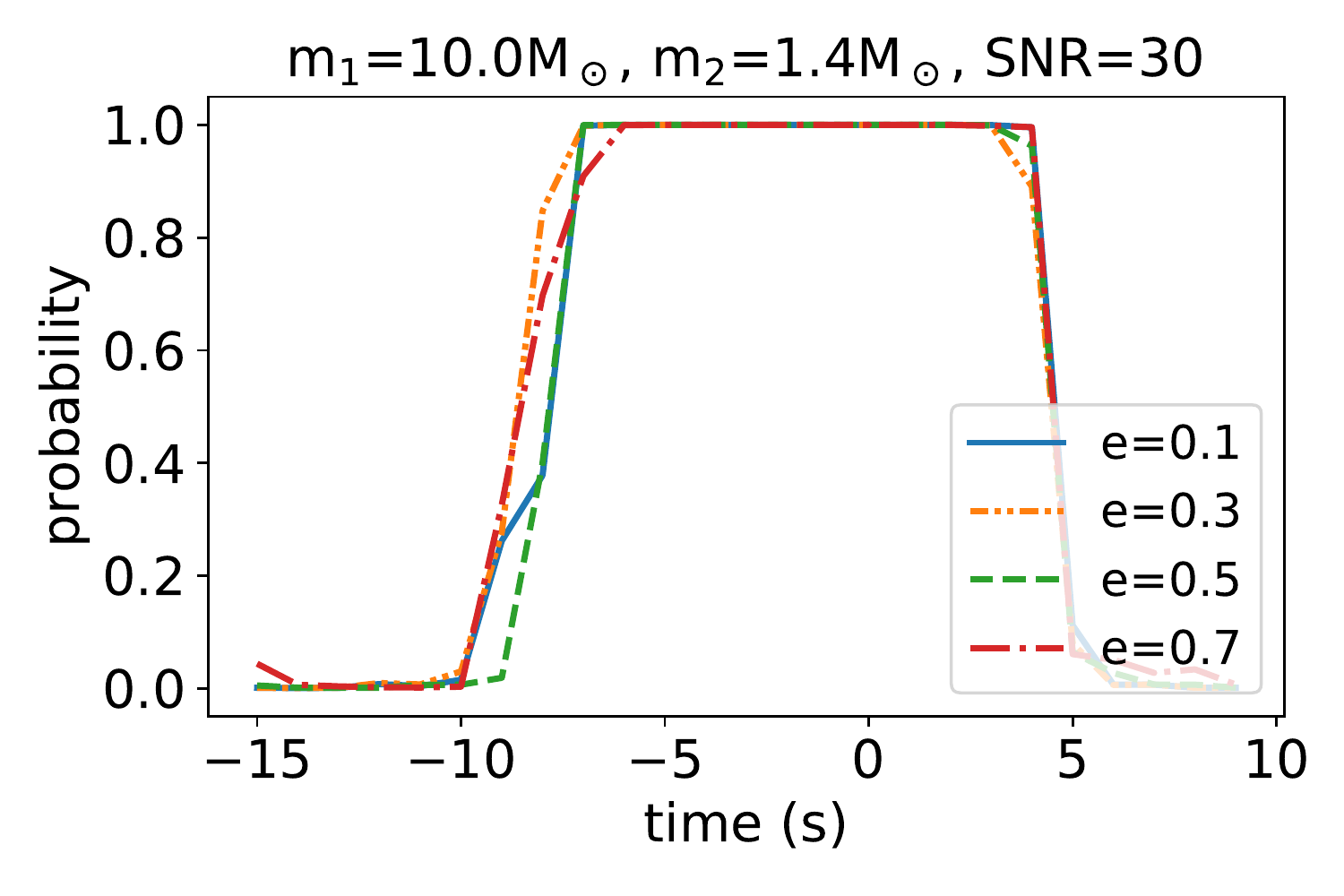}
}
\centerline{
\includegraphics[width=0.45\linewidth]{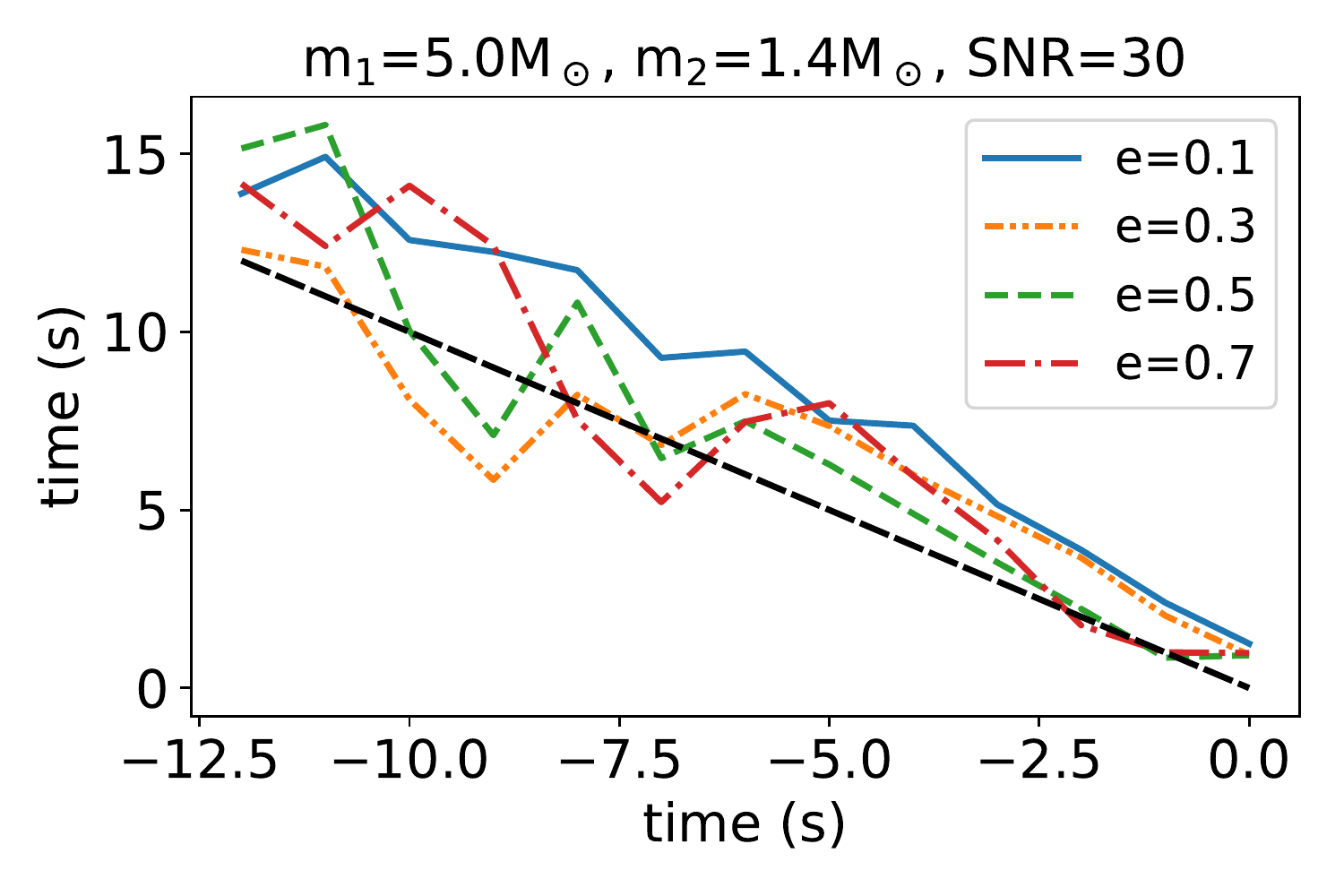}
\includegraphics[width=0.45\linewidth]{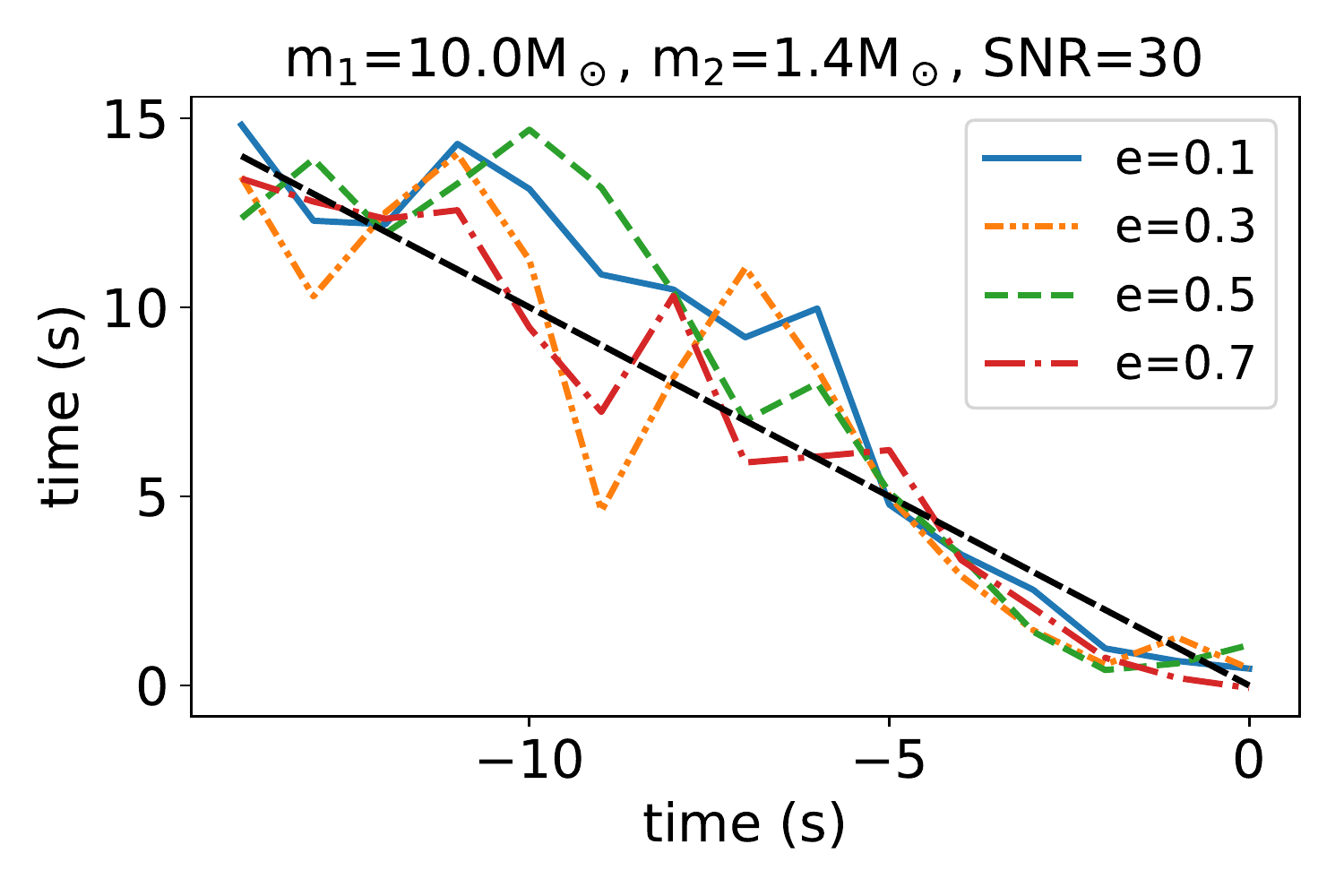}
}
\centerline{
\includegraphics[width=0.45\linewidth]{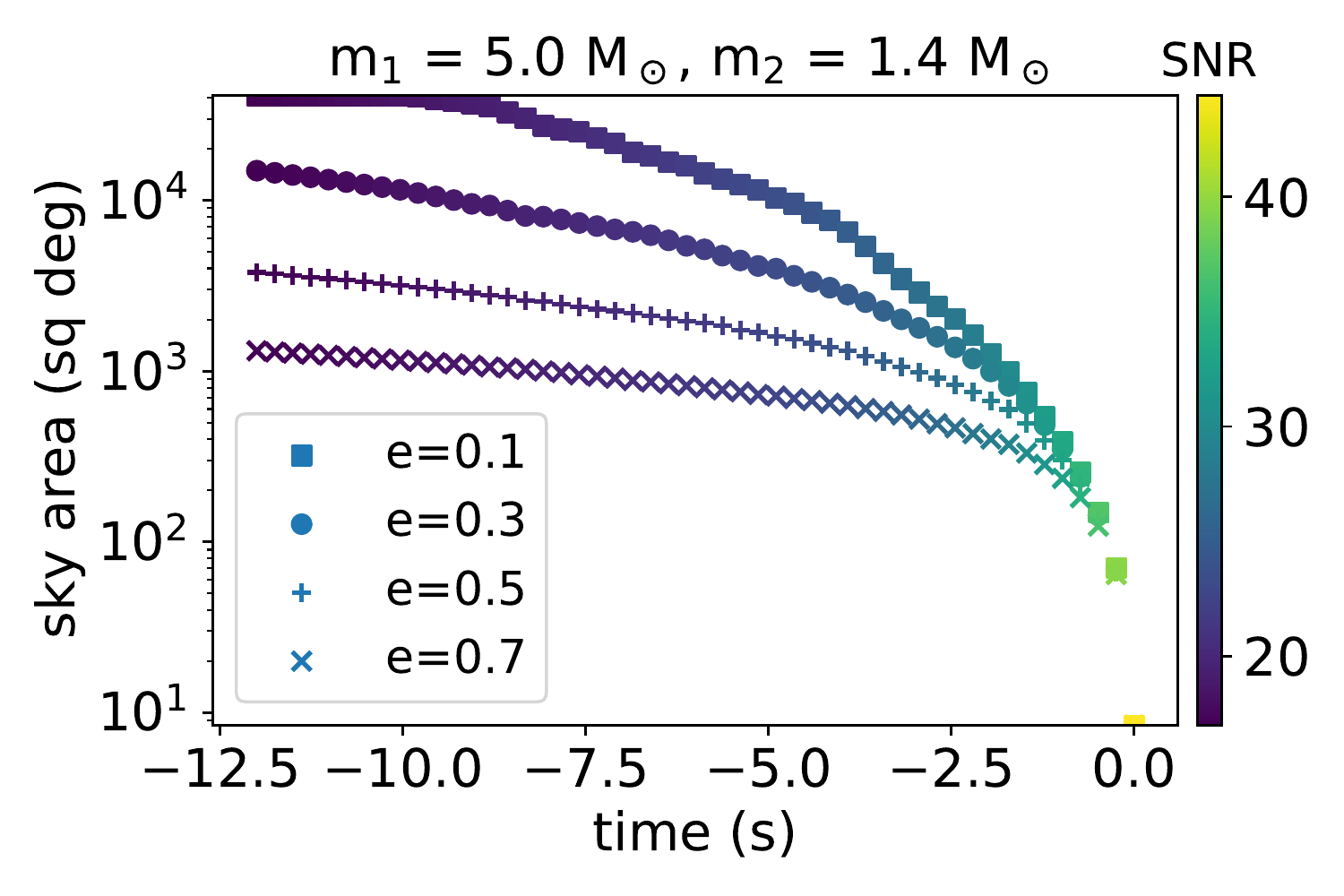}
\includegraphics[width=0.45\linewidth]{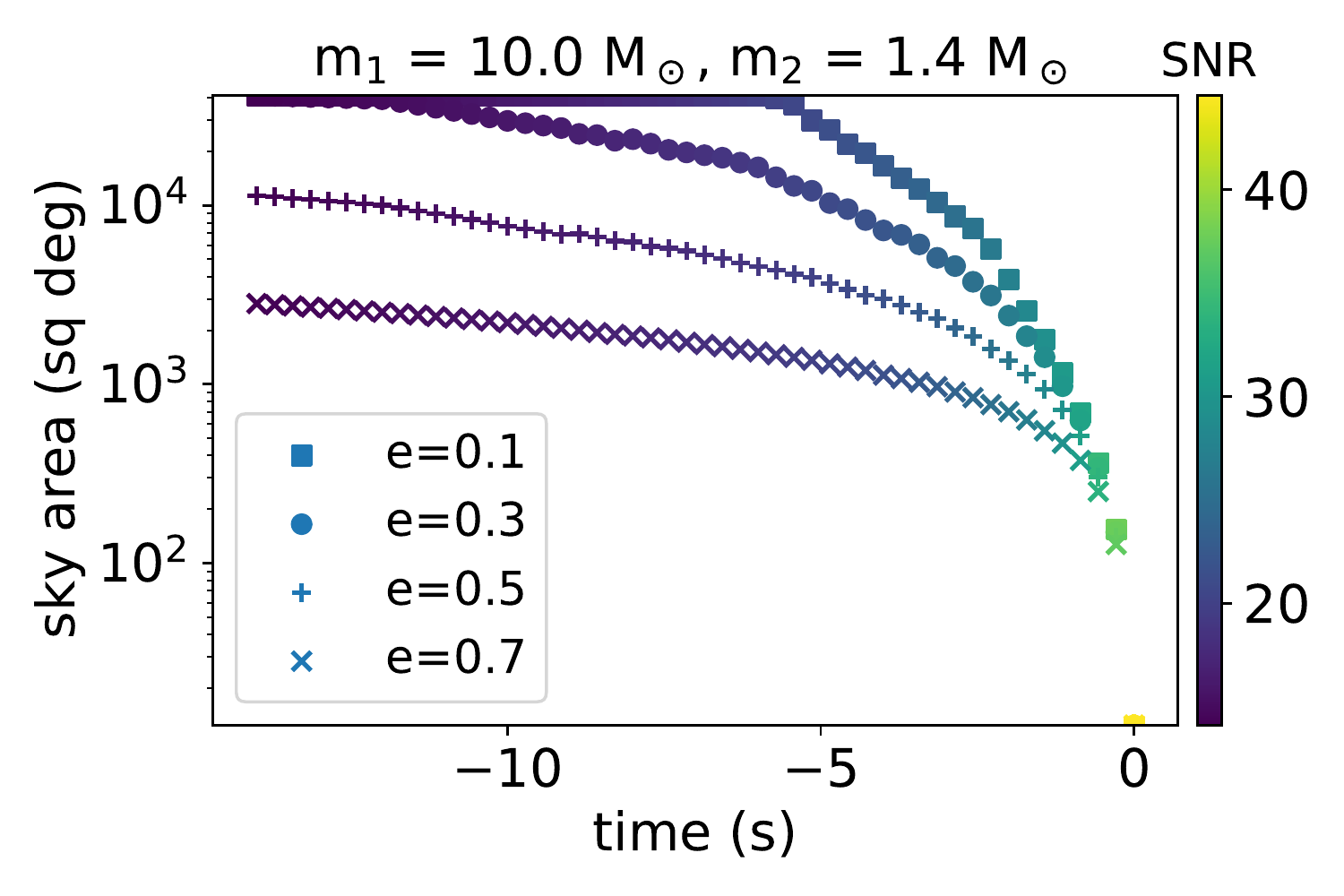}
}
\caption{\textbf{Top panels} Neural networks identify injections of neutron star-black hole waveforms in O2 LIGO data up to 12 seconds before merger. \textbf{Mid panels} Neural networks provide reliable estimates of time to merger 5 seconds ahead of the actual merger event.  \textbf{Bottom panels} Time-dependent sky localization of injected neutron star-black hole waveforms in O2 LIGO noise.}
\label{fig:ecc_nsbh}
\end{figure*}

Figure~\ref{fig:ecc_nsbh} summarizes our findings for injections 
that describe binaries with component masses 
\((m_1=5\msun, m_2=1.4\msun)\) and \((m_1=10\msun, m_2=1.4\msun)\) 
and \(\textrm{SNR}=30\). The top panels in this figure 
show that forecasting is weekly 
dependent on eccentricity, and that systems with lower total mass 
may be identified up to 12 seconds before merger, whereas heavier 
systems are identified 10 seconds before merger. This behaviour 
is expected due to several factors. 
First, forecasting results are optimal for low mass 
black hole-neutron star systems because, as shown in the 
left panel of Figure~\ref{fig:ecc_spectrogram}, the 
time evolution of the whitened waveform amplitude undergoes a 
gradual increase as it nears merger. On the other hand, 
this evolution becomes more asymmetrical, characterized by 
a sharp amplitude growth near merger, as we consider 
heavier black holes. As a result, the spectrograms used 
to identify the existence of waveforms in advanced LIGO noise 
contain a wealth of information near merger where the power 
is concentrated. In turn, neural nets become more 
confident of the existence of these systems closer to merger. 

The mid panels in Figure~\ref{fig:ecc_nsbh} 
show that neural networks may provide reliable information 
about the merger time about 6 seconds before merger. The 
bottom panels show that, as in the 
case of binary neutron stars, sky localization 
depends strongly on orbital eccentricity, and that the most 
eccentric systems may be better localized before merger. From the 
time our deep learning algorithms identify these injections, 
their sky localization is reduced from 
\(\sim O(10^3)\) square degrees for the most eccentric systems and 
\(\sim O(10^4)\) square degrees for the least eccentric systems, 
to only \(\sim O(10)\) square degrees at merger. It is worth noting 
that this improvement takes place within 10 seconds.

\subsection{Eccentric Binary Black Hole Mergers}
\label{sec:bbh}

The third set of systems we considered are 
binary black hole mergers with component masses 
\(m_{\textrm{BH}}\in[3\msun, 15\msun]\) and 
eccentricities \(e\leq0.9\). As in the two previous cases, 
we used injections that sample a wide range of 
inclination angles, sky locations and SNRs.

\begin{figure*}
\centerline{
\includegraphics[width=0.45\linewidth]{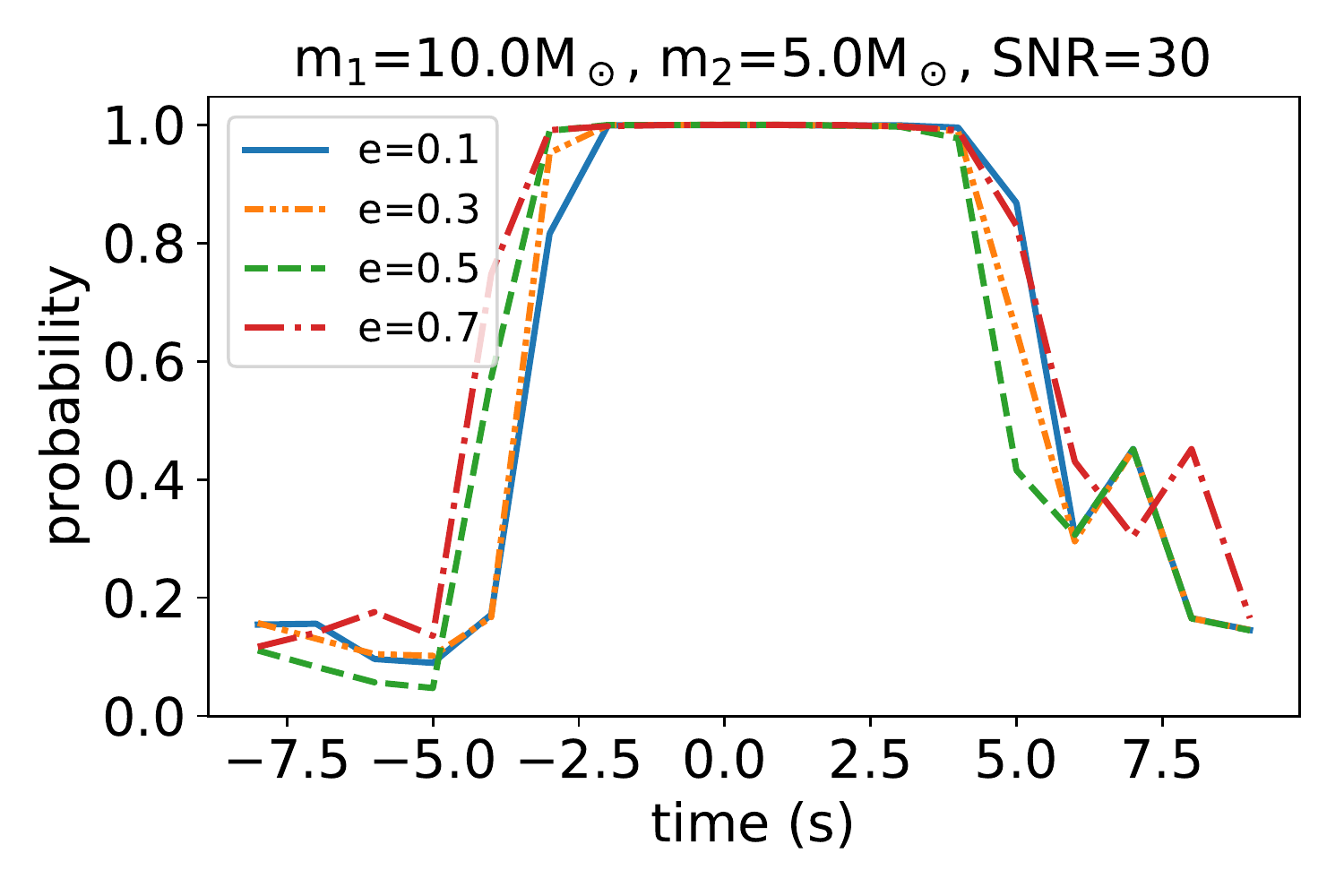}
\includegraphics[width=0.45\linewidth]{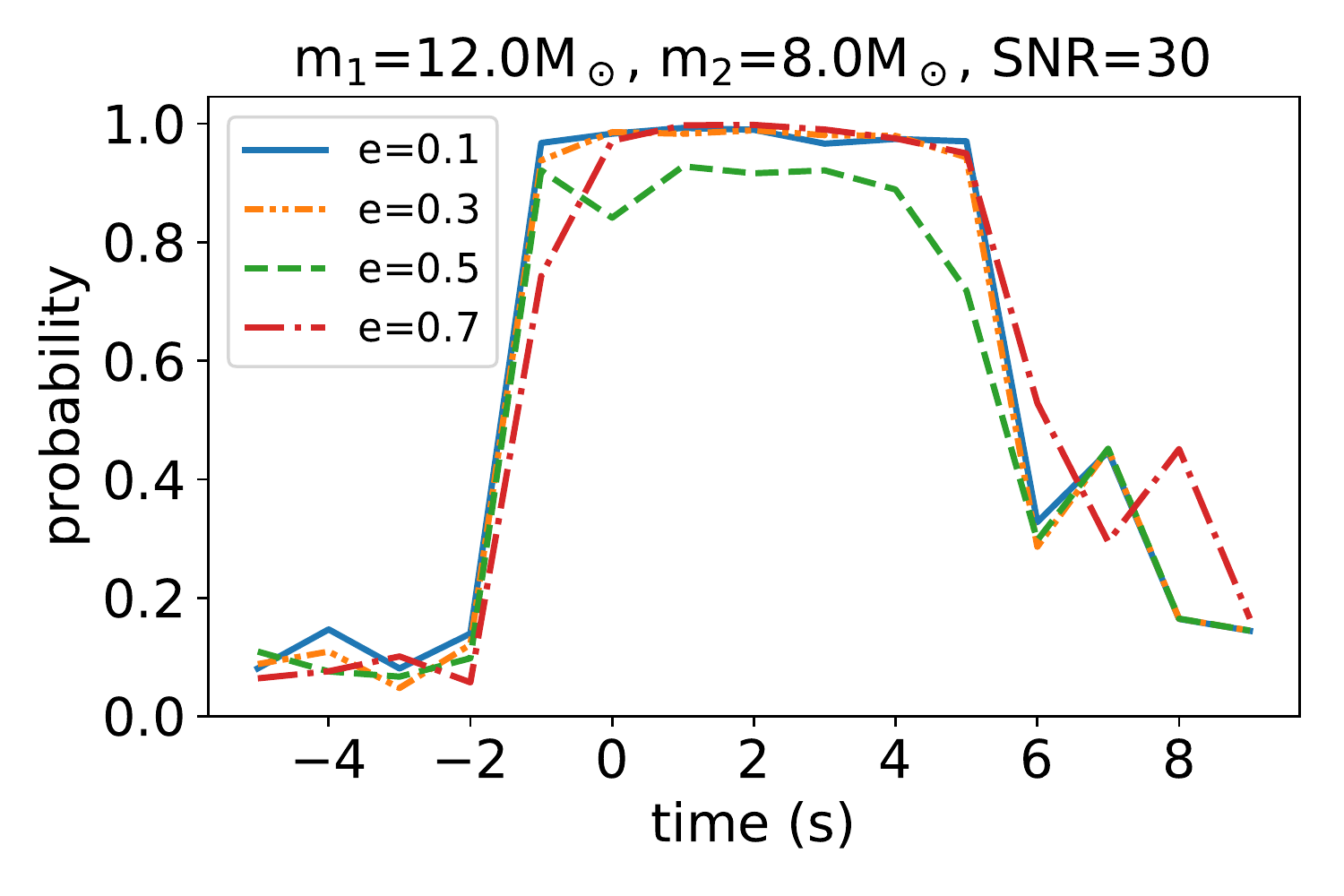}
}
\centerline{
\includegraphics[width=0.45\linewidth]{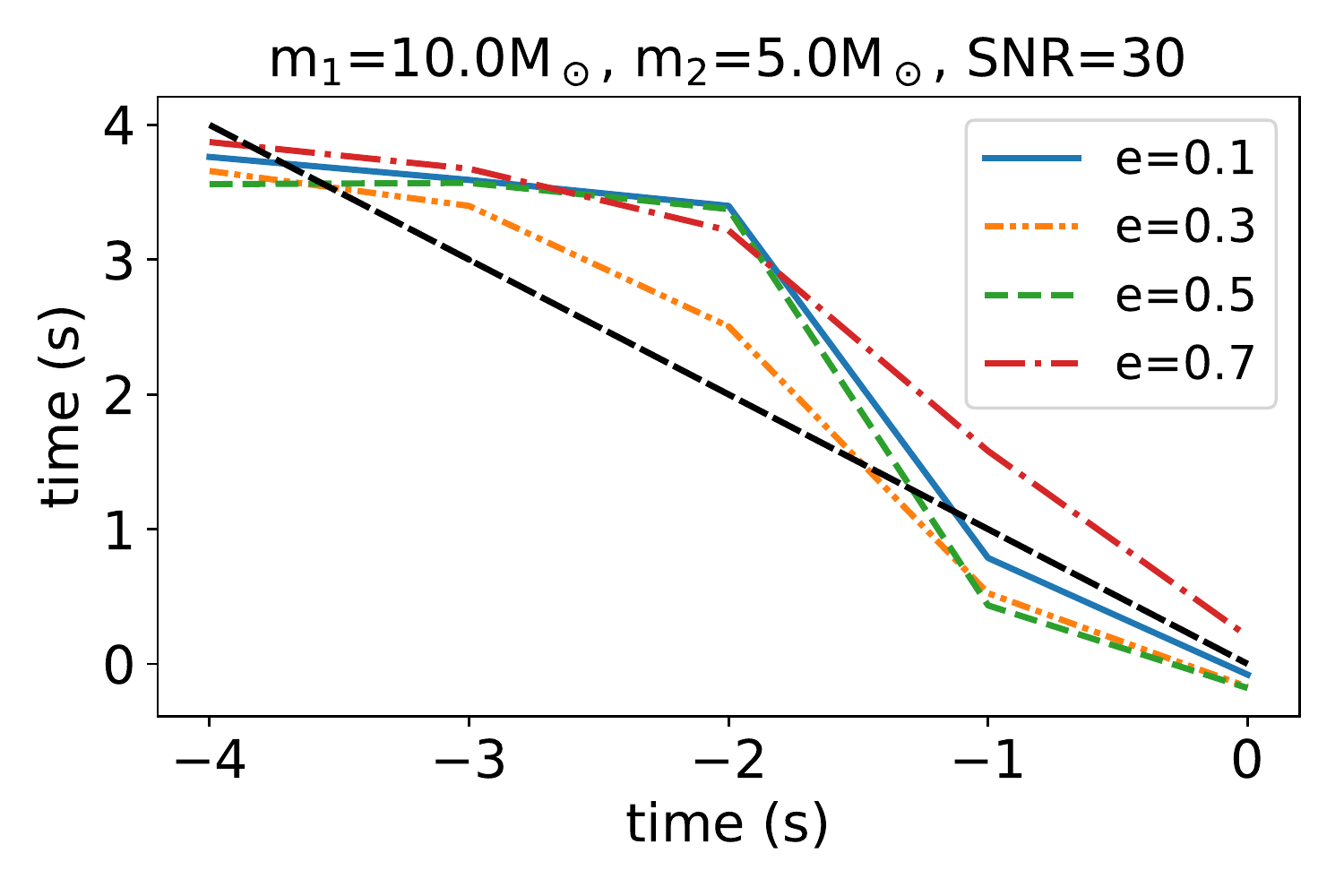}
\includegraphics[width=0.45\linewidth]{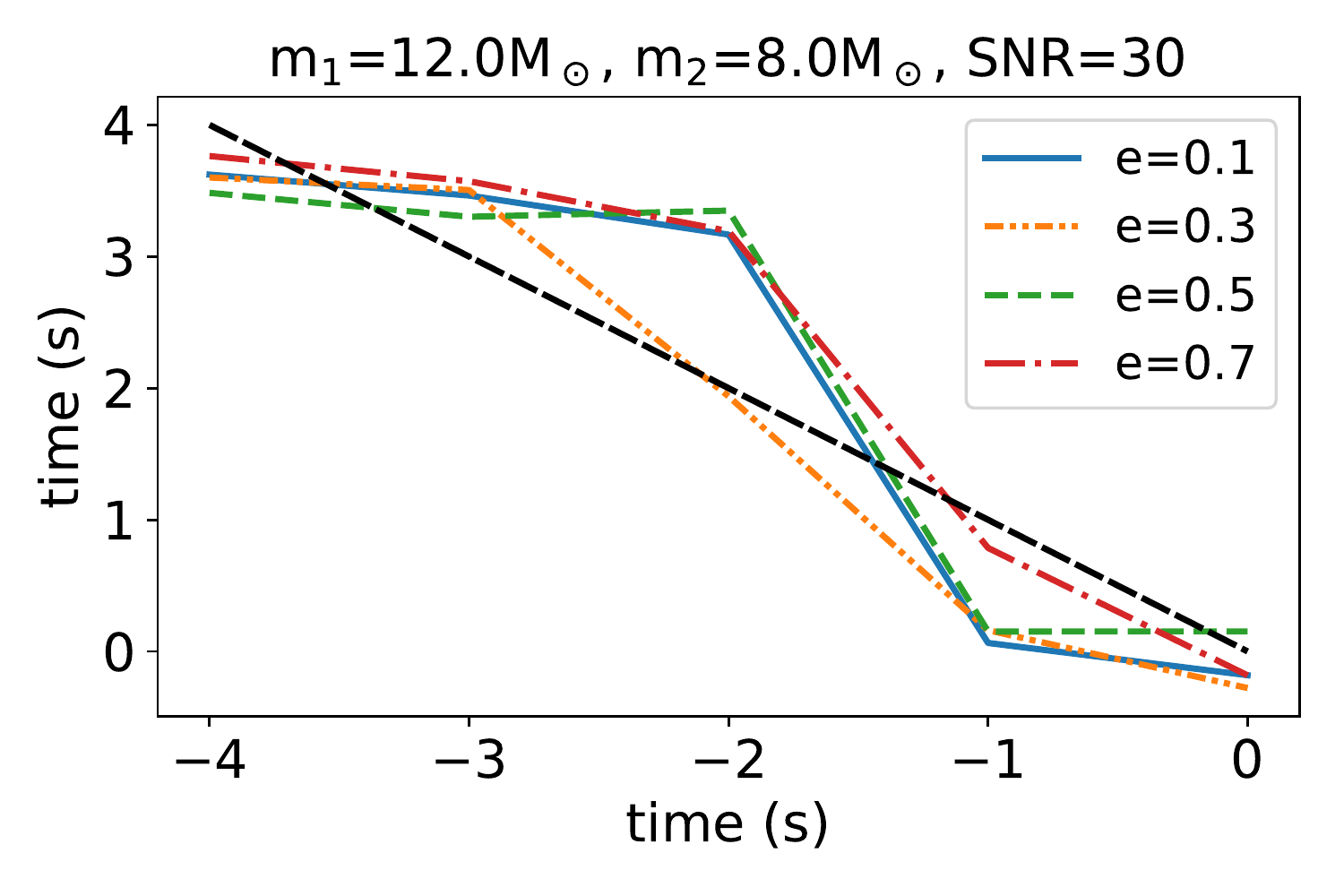}
}
\centerline{
\includegraphics[width=0.45\linewidth]{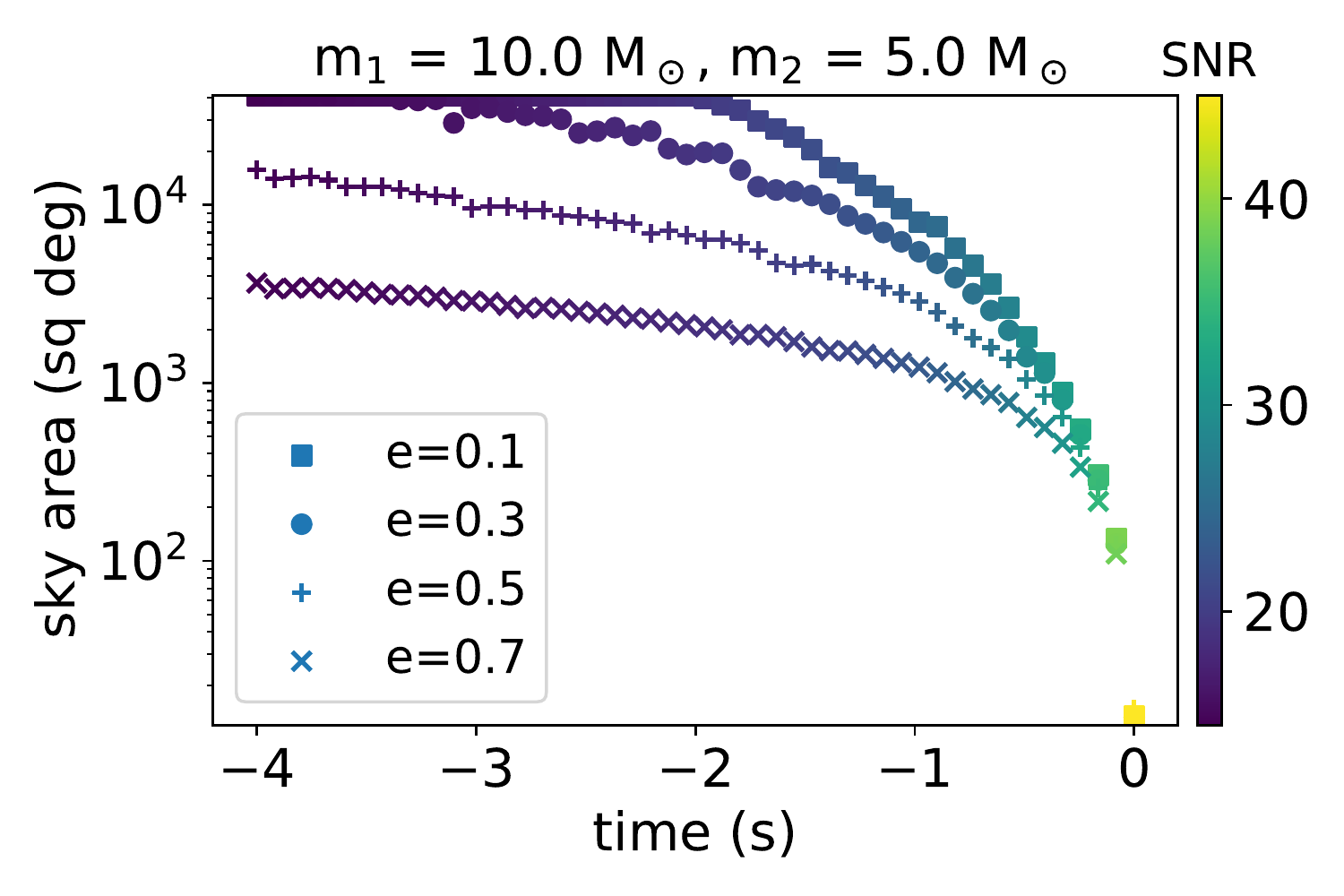}
\includegraphics[width=0.45\linewidth]{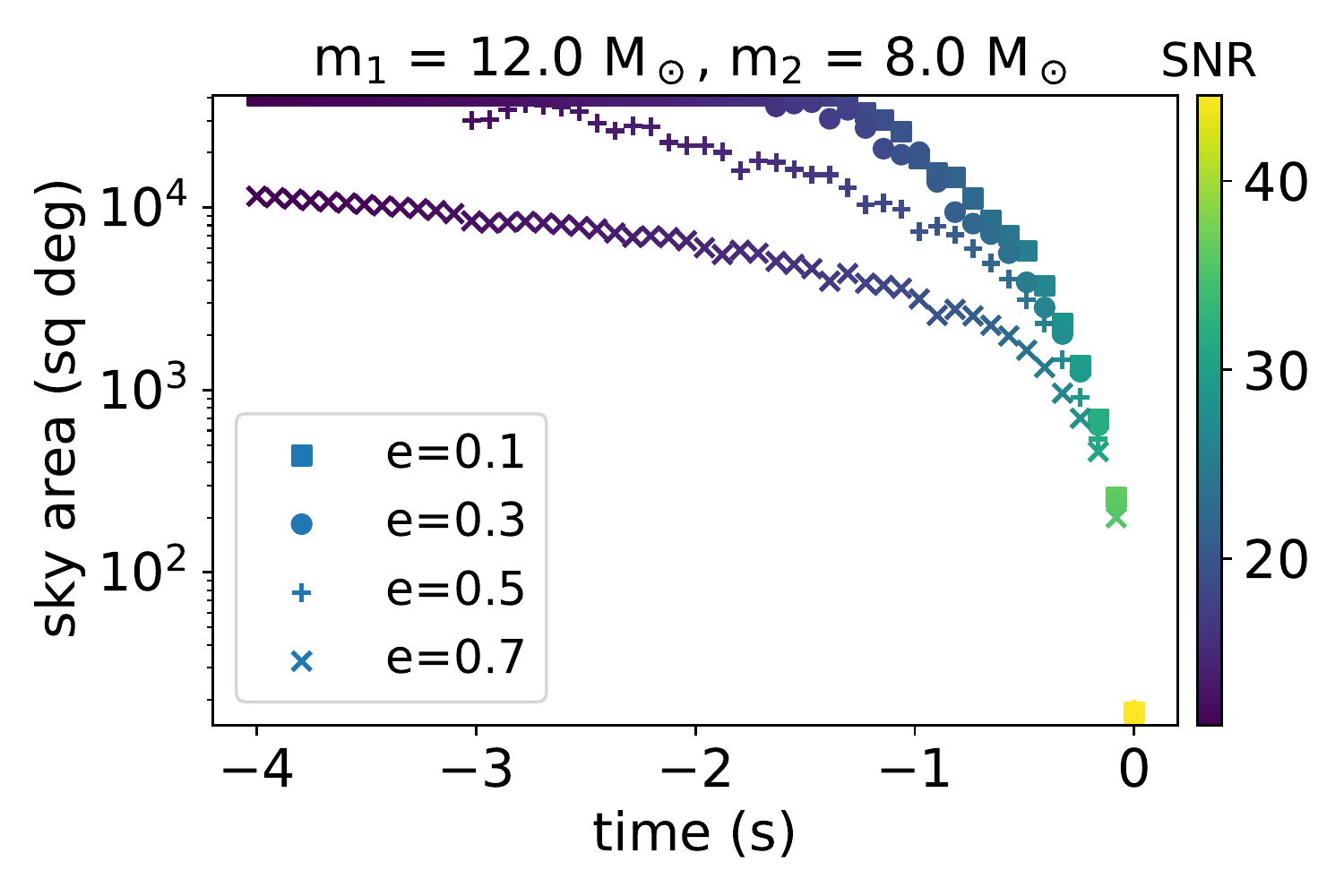}
}
\caption{\textbf{Top panels} Neural networks identify injections of binary black hole waveforms in O2 LIGO data a few seconds before merger. Forecasting results transition into real-time alerts for these black hole mergers.  \textbf{Mid panels} Neural networks provide a reliable estimate of the merger once the coalescence is imminent.  \textbf{Bottom panels} Time-dependent sky localization of injected binary black holes in O2 LIGO noise.}
\label{fig:ecc_bbh}
\end{figure*}

Our findings are summarized in Figure~\ref{fig:ecc_bbh} for 
two sample systems with component masses 
\((m_1=10\msun, m_2=5\msun)\) and \((m_1=12\msun, m_2=8\msun)\) 
and \(\textrm{SNR}=30\). 

The top panels in Figure~\ref{fig:ecc_bbh} show that deep learning 
may forecast the merger of low mass binary black hole 
mergers about 2 seconds before merger. The mid panels indicate 
that deep learning may indicate the time to merger about 
1 second before the binary components collide. In other words, 
our deep learning algorithms are actually working as real-time 
gravitational wave classifiers. The bottom panels show 
how accurately we may constrain the sky location of binary black 
hole mergers. These results show that we may localize these 
sources within \(\sim O(10^2)\) square degrees in the vicinity of 
merger merger, and down to \(\sim O(10)\) square degrees at 
merger. 

\section{Quantized neural networks for rapid, energy efficient forecasting}
\label{sec:quant}

We compare the results of inference speed and error rate for 
un-quantized and quantized networks trained with spectrogram images. Prior to quantization, the networks need to be calibrated using 
a set of testing images. We randomly select 30 images from the 
testing data set to yield the optimal scales and zero points 
for activation tensors. We selected 1000 spectrogram images from 
the testing set to benchmark the performance. After quantization, 
the inference latency of the quantized model is 8 ms per prediction, while that of the un-quantized model is 20 ms per prediction, 
showing a 2.5x speedup. Here we used a batch number of 16, 
which means the network predicted 16 images at a time to exploit the parallelism of the multi-core processor. By converting 
the model parameters from floating-point to fixed-point representation, we can reduce the overall model size by 4x, 
from 92 MB to 23 MB. Our results show that quantization does 
not negatively affect inference performance, and, 
in our case, it is able to decrease the top-1 error 
rates by 10\%. 

\begin{figure*}[!htp]
\centerline{
\includegraphics[width=0.45\linewidth]{fig/bns_1.pdf}
\includegraphics[width=0.45\linewidth]{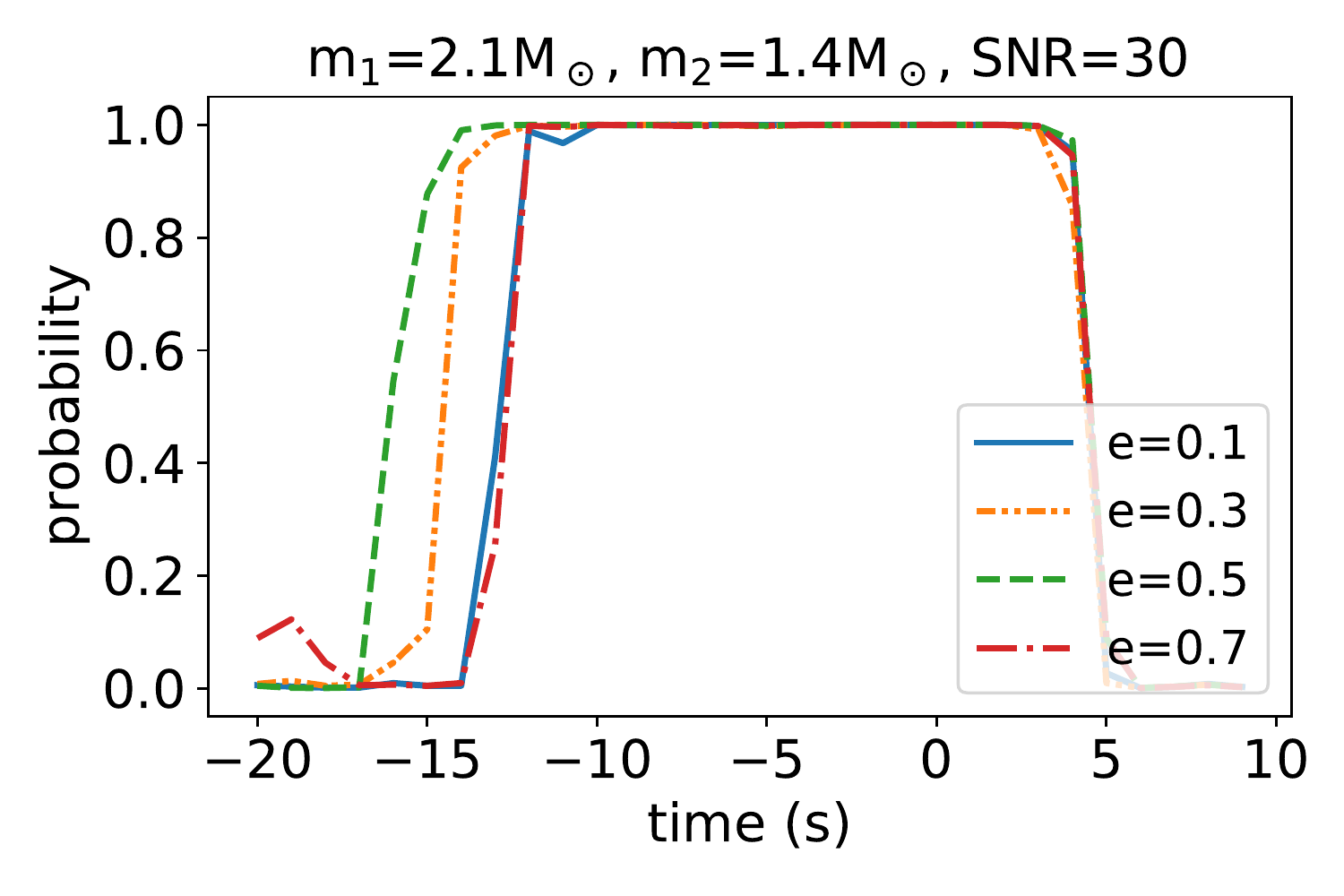}
}
\centerline{
\includegraphics[width=0.45\linewidth]{fig/nsbh_0.pdf}
\includegraphics[width=0.45\linewidth]{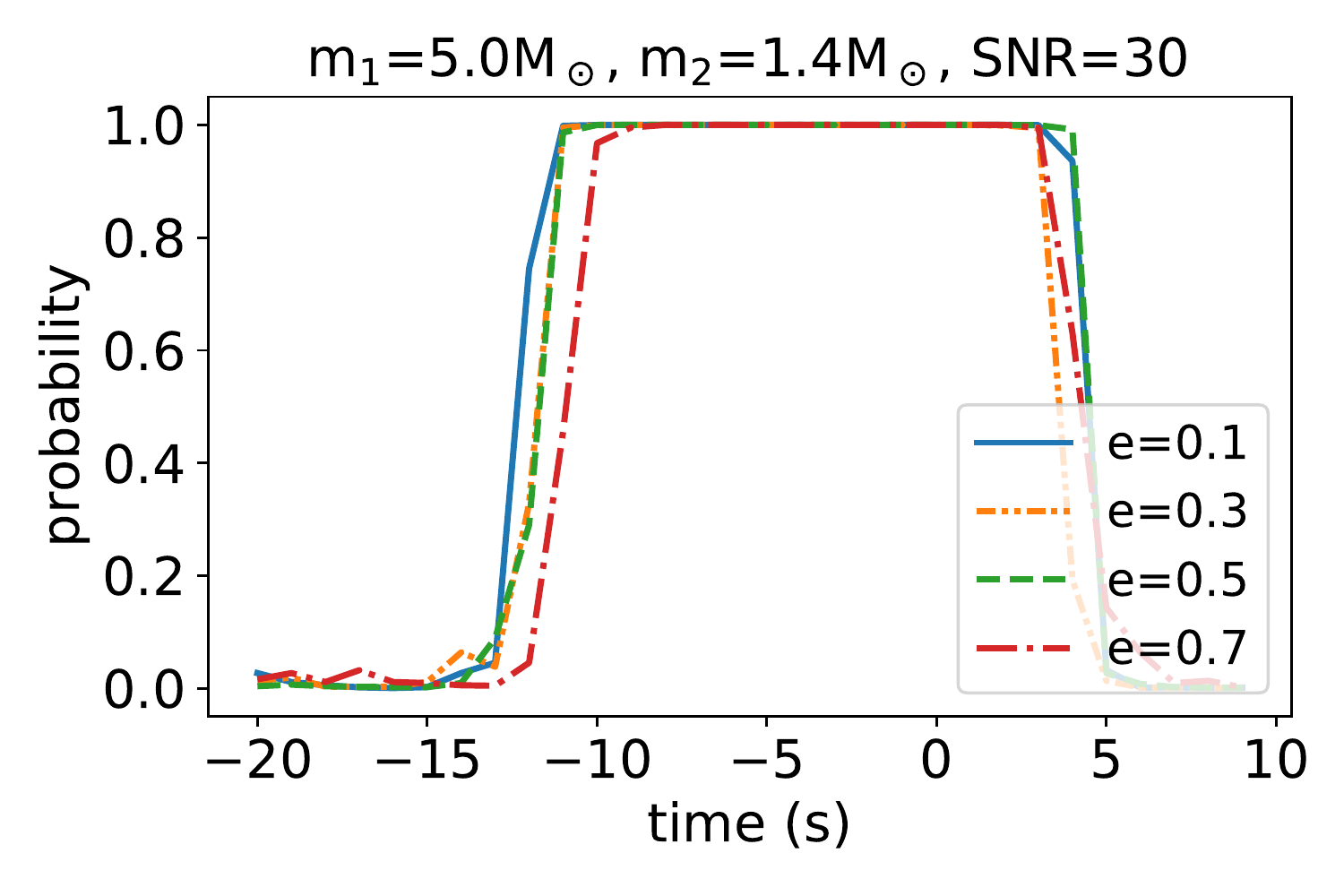}
}
\centerline{
\includegraphics[width=0.45\linewidth]{fig/bbh_0.pdf}
\includegraphics[width=0.45\linewidth]{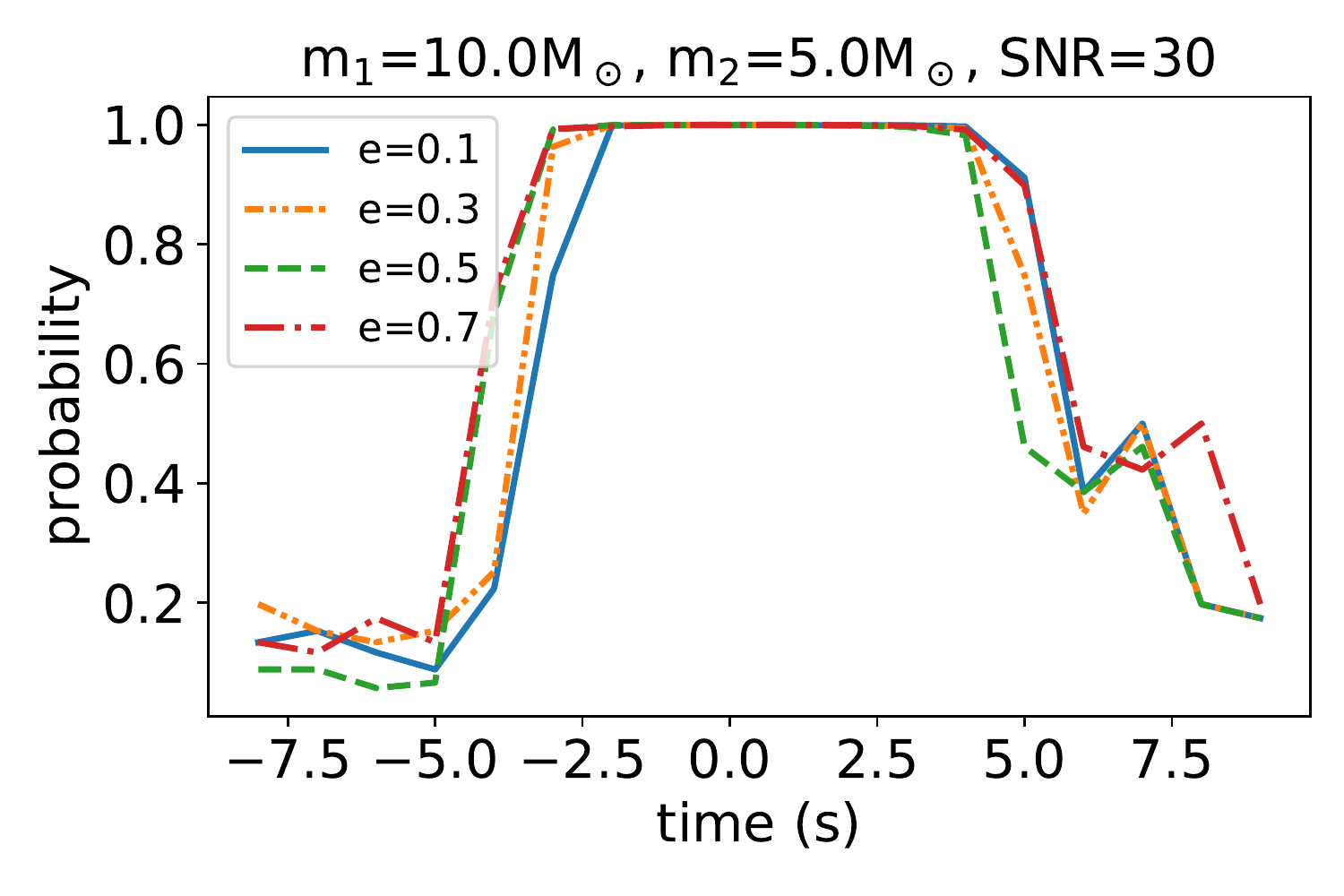}
}
\caption{Deep learning forecasting for binary  neutron  stars, black hole-neutron star systems, and  binary black hole mergers in advanced LIGO data, from top to bottom respectively. The left panel shows results produced by un-quantized networks, and the right panel shows results produced by quantized networks.}
\label{fig:ecc_quant}
\end{figure*}

Figure~\ref{fig:ecc_quant} presents three forecasting results produced by our un-quantized and quantized networks. We select one scenario from each of these three sources: binary  neutron  stars, black hole-neutron star systems, and binary black hole mergers. As illustrated, the forecasting results produced by the quantized network are very similar to those produced by the un-quantized network, but faster. Furthermore, our quantized networks alleviate the high demand for computational and memory resources, and, as a result, the power efficiency is increased for gravitational wave forecasting, which is critical for edge devices.

\section{Conclusions}
\label{sec:end}

We have introduced the first application of deep 
learning to forecast the merger of eccentric binary systems 
in advanced LIGO noise. The neural networks introduced 
in this study can readily identify eccentric waveforms 
injected in O2 noise up to 15 seconds before merger, 
and estimate the time within 
which the binary components will coalesce. We have 
adapted a sky localization method to estimate the time-dependent 
sky area within which we may observe these signals, from 
the time they are identified by our forecasting algorithms 
up to merger, with a three detector network 
encompassing the twin LIGO detectors (Hanford and Livingston) 
and the Virgo detector. These results were obtained using 
open source LIGO and Virgo 
O2 noise. Our findings indicate that the performance 
of our forecasting neural networks is fairly independent of 
the eccentricity of the binary system under consideration. 
However, sky localization improves significantly for 
larger values of eccentricity.

We quantized our neural networks and found that they improve 
the latency and power efficiency of gravitational wave forecasting, making it suitable for edge computing. These algorithms may then 
enable a broader cross section 
of the community to readily use these algorithms for 
time-sensitive multi-messenger astrophysics discovery campaigns 
that target compact binary systems that may reside in dense stellar 
environments, but that due to their complex morphology 
are difficult to capture with other signal processing tools. 

\section{Acknowledgments}
\label{ack}

We thank Frans Pretorius for insightful conversations 
that led to the conceptualization of this project. We 
gratefully acknowledge National Science Foundation (NSF) awards 
OAC-1931561 and OAC-1934757. We thank NVIDIA for their 
continued support. This work utilized resources supported 
by the NSF's Major Research Instrumentation 
program, the Hardware-Learning Accelerated (HAL) cluster, 
grant OAC-1725729, as well as the University of Illinois 
at Urbana-Champaign. N.L. acknowledges support from NSF 
grant PHY-1912171, the Simons Foundation, and the Canadian 
Institute for Advanced Research (CIFAR)

\clearpage

\appendix
\restartappendixnumbering
\section{Waveforms used for neural network training}

We have mentioned in previous studies that deep learning 
enables the generalization to new types of signals, beyond 
the waveform set used for training. We have explored 
this assertion by comparing forecasting 
predictions for the binary neutron star system GW170817 
using neural networks trained with the eccentric 
waveform model used in this analysis~\citep{East:2012xq}, and the 
\texttt{IMRPhenomD\_NRTidal}~\citep{Dietrich:2018uni} 
waveform model . 

\begin{figure*}[!htp]
\centerline{
\includegraphics[width=0.5\linewidth]{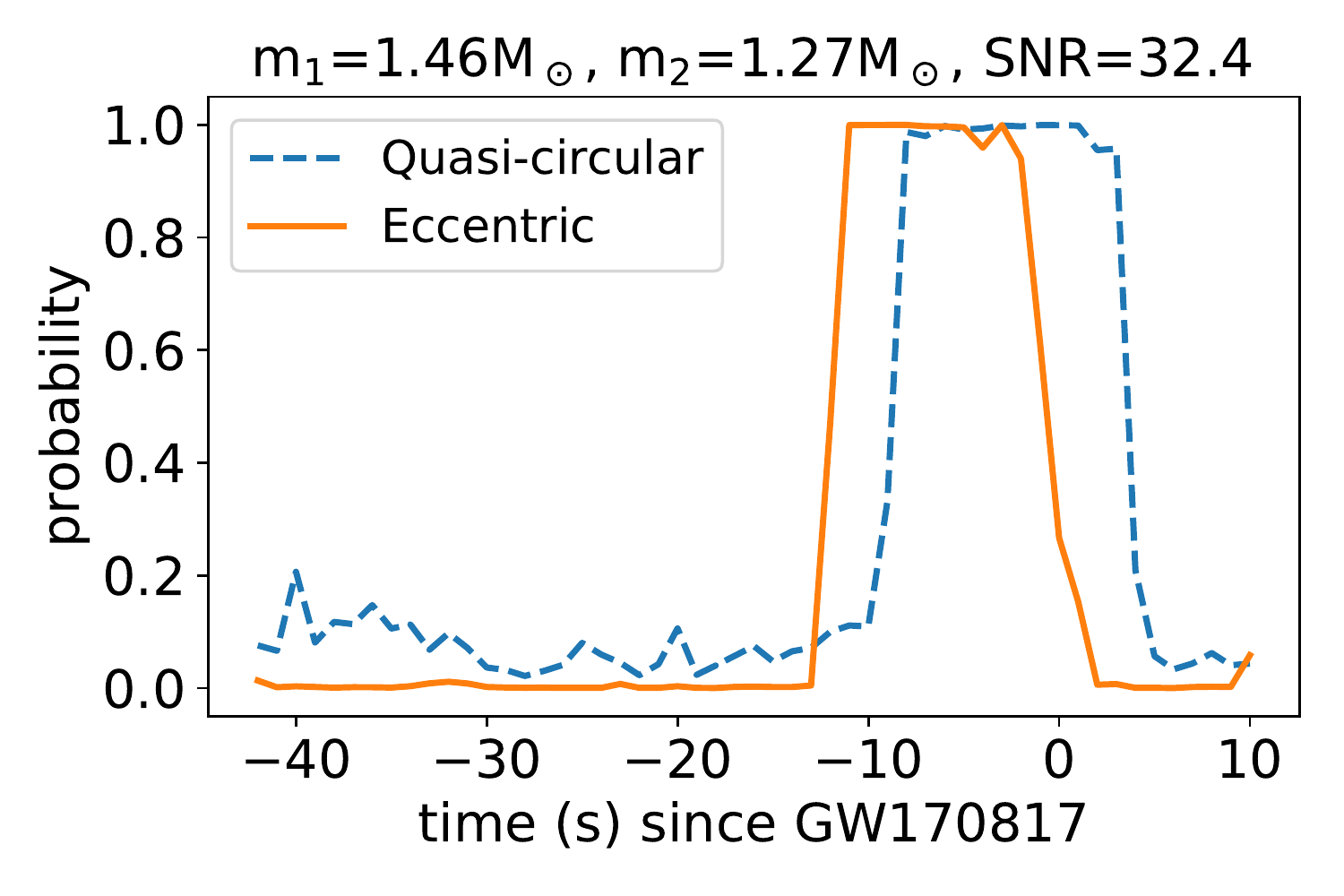}
}
\caption{Forecasting of the binary neutron star GW170817 with a neural network trained with \texttt{IMRPhenomD\_NRTidal} waveforms (Quasi-circular) and with eccentric waveforms produced by the model introduced in~\citep{East:2012xq} (Eccentric).}
\label{fig:gw170817}
\end{figure*}

Figure~\ref{fig:gw170817} indicates that both neural network 
models have similar forecasting capabilities~\citep{Wei_forecast}. 
These results indicate 
that while we should continue to use the best waveform models 
available to train deep learning algorithms, neural networks 
may also be used to enable data-driven discovery by guiding them 
towards the right answer with approximate models that describe 
complex physical processes. 

\section{Gravitational wave forecasting for the gravitational wave events GW190814 and GW190412.}

\begin{figure*}[!htp]
    \centerline{
    \includegraphics[width=0.45\linewidth]{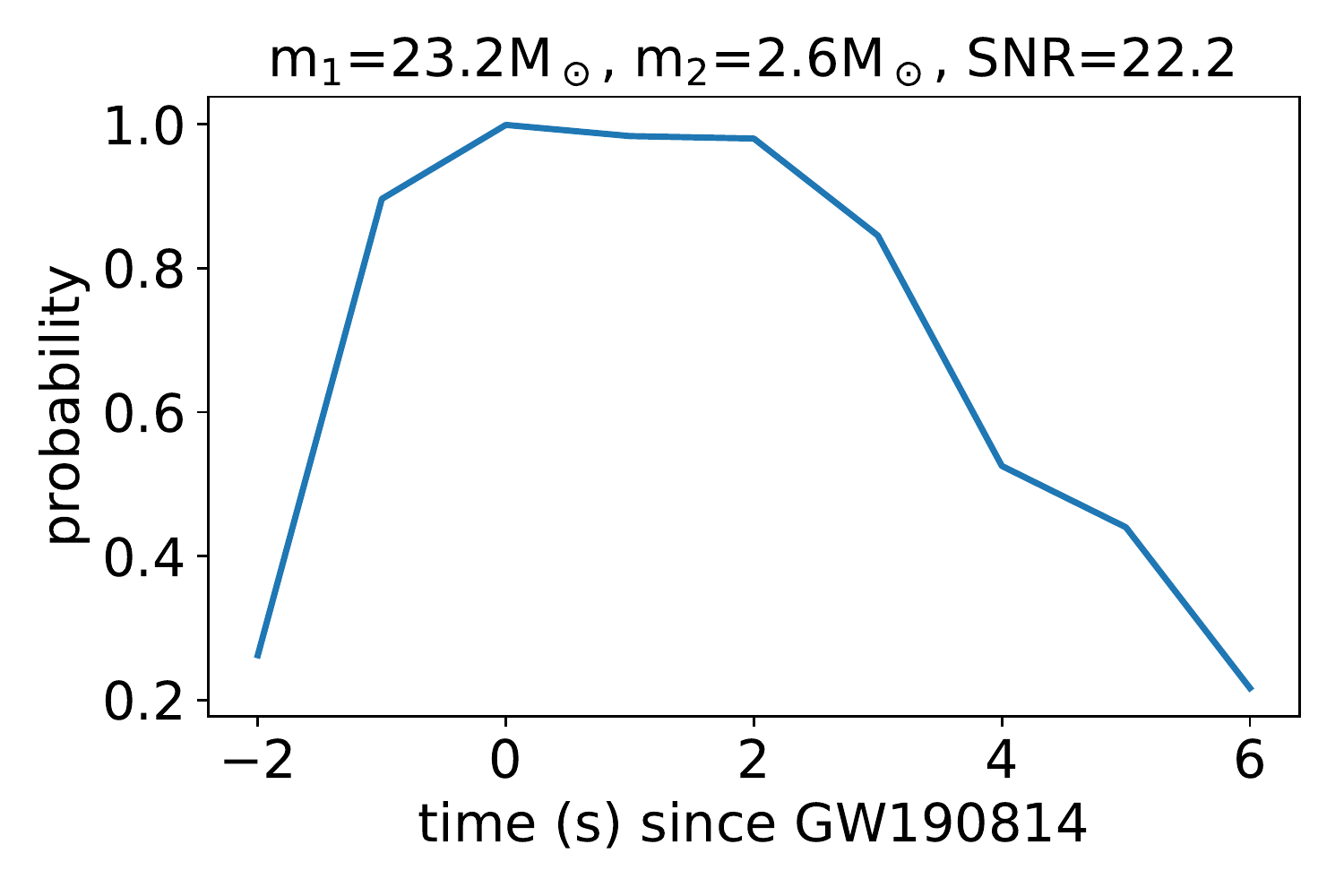}
    \includegraphics[width=0.45\linewidth]{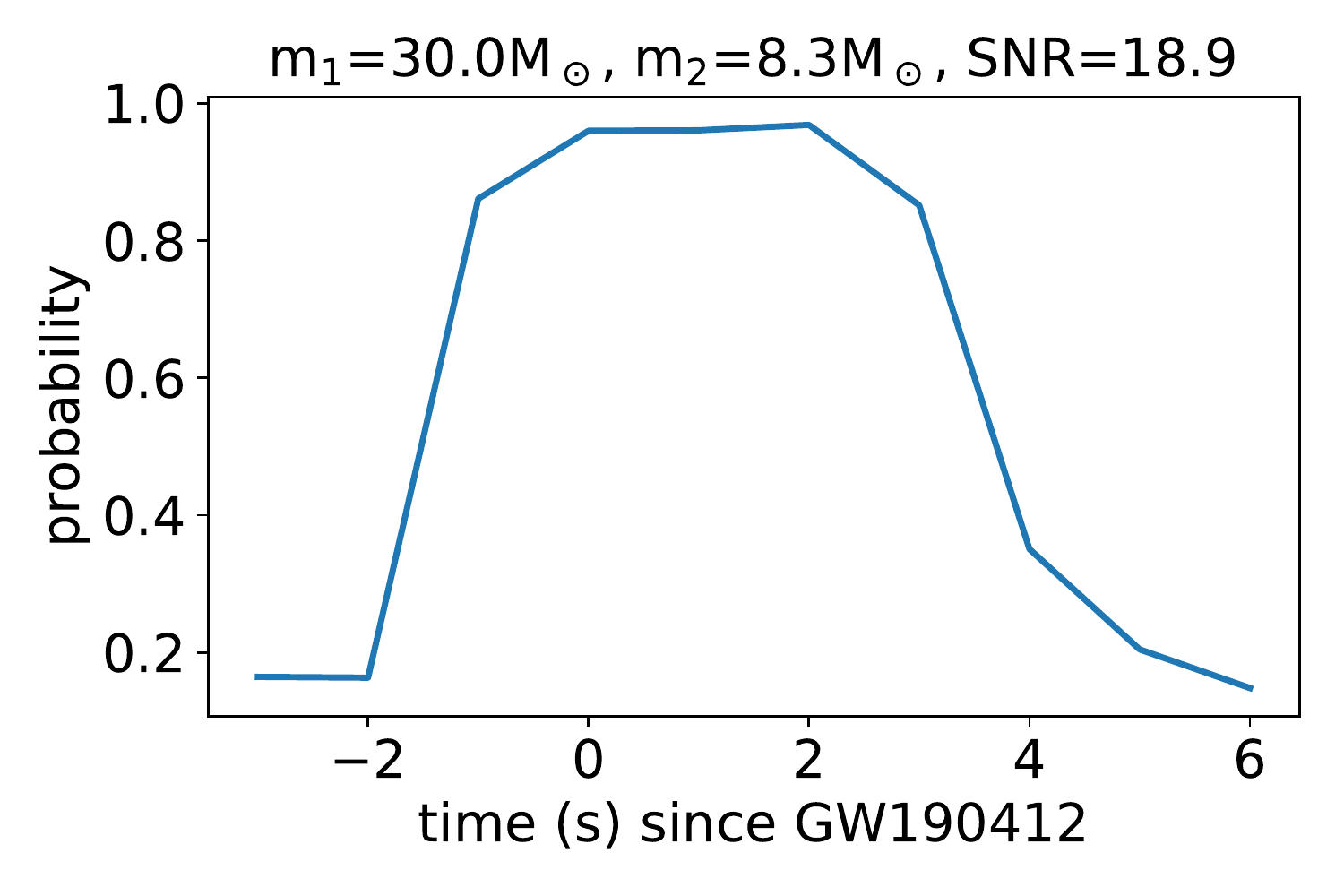}
    }
    \caption{Application of our neural networks to forecast the merger of the events GW190814 and GW190412.}
    \label{fig:gw190814}
\end{figure*}

Figure~\ref{fig:gw190814} presents forecasting results 
for the events GW190814 and GW190412. These results are 
consistent with short early warning or real-time detection 
alerts that we discussed in 
Sections~\ref{sec:nsbh} and~\ref{sec:bbh}. The key point here is that 
the larger the total mass of the systems under consideration 
the closer to the merger event our forecasting algorithms identify 
gravitational wave signals.

\clearpage

\bibliography{awareness}
\bibliographystyle{aasjournal}

\end{document}